\documentclass[10pt,journal,compsoc]{IEEEtran}
\usepackage{amssymb}
\usepackage{amsbsy}
\usepackage{footnote}
\usepackage{amsmath}
\usepackage{enumitem}
\usepackage{array}
\usepackage{bbding}
\usepackage{amsfonts}
\makeatletter
\let\NAT@parse\undefined
\makeatother
\usepackage[colorlinks,linkcolor=blue,citecolor=blue]{hyperref}
\usepackage[usenames,dvipsnames]{color}
\usepackage{colortbl}
\usepackage{multicol}
\usepackage{multirow}
\usepackage{booktabs}
\usepackage{makecell}
\usepackage{amsthm}
\definecolor{gray1}{gray}{.3}
\definecolor{gray2}{gray}{.5}
\definecolor{gray3}{gray}{.7}

\newtheorem{definition}{Definition}
\newtheorem{theorem}{Theorem}
\ifCLASSOPTIONcompsoc
  \usepackage[nocompress]{cite}
\else
  \usepackage{cite}
\fi
\ifCLASSINFOpdf
   \usepackage[pdftex]{graphicx}
   \usepackage{subfigure}
\else
\fi
\begin{document}
%
\title{Representation Learning on Heterostructures via Heterogeneous Anonymous Walks}
%
%
%
%

\author{
       Xuan~Guo,
       Pengfei~Jiao,
       Ting~Pan,
       Wang~Zhang,
       Mengyu Jia,
       Danyang Shi
       and Wenjun Wang*


\thanks{X. Guo,
       T. Pan,
       W. Zhang,
       M. Jia,
       D. shi
       and W. Wang are with the College of Intelligence and Computing, Tianjin University, Tianjin, 300350, China (email: \{guoxuan, tingpan, wangzhang, myjia, shidanyang, wjwang\}@tju.edu.cn).}
\thanks{P. Jiao is with the School of Cyberspace, Hangzhou Dianzi University, Hangzhou, 310018, China (email: pjiao@hdu.edu.cn).}
\thanks{* denotes the corresponding author.}
 
}
%
%


\markboth{Journal of \LaTeX\ Class Files,~Vol.~14, No.~8, August~2015}%
{Shell \MakeLowercase{\textit{et al.}}: Bare Demo of IEEEtran.cls for Computer Society Journals}

%



\IEEEtitleabstractindextext{%
\begin{abstract}
Capturing structural similarity has been a hot topic in the field of network embedding recently due to its great help in understanding the node functions and behaviors. However, existing works have paid very much attention to learning structures on homogeneous networks while the related study on heterogeneous networks is still a void. In this paper, we try to take the first step for representation learning on heterostructures, which is very challenging due to their highly diverse combinations of node types and underlying structures. To effectively distinguish diverse heterostructures, 
we firstly propose a theoretically guaranteed technique called heterogeneous anonymous walk (HAW) and its variant coarse HAW (CHAW). Then, we devise the heterogeneous anonymous walk embedding (HAWE) and its variant coarse HAWE in a data-driven manner to circumvent using an extremely large number of possible walks and train embeddings by predicting occurring walks in the neighborhood of each node. Finally, we design and apply extensive and illustrative experiments on synthetic and real-world networks to build a benchmark on heterostructure learning and evaluate the effectiveness of our methods. The results demonstrate our methods achieve outstanding performance compared with both homogeneous and heterogeneous classic methods, and can be applied on large-scale networks.
\end{abstract}

\begin{IEEEkeywords}
Network Embedding, Heterogeneous Network, Structural Similarity, Role Discovery, Unsupervised Learning.
\end{IEEEkeywords}}

\maketitle

\IEEEdisplaynontitleabstractindextext

%
\IEEEpeerreviewmaketitle
\section{Introduction}
\label{sec-introduction}

Network Embedding (NE) \cite{zhang2018network,boguna2021network} has been a rolling network science and representation learning bandwagon in recent years. Researchers' enthusiasm for NE stems from its ability to transform large-scale unstructured data into low-dimensional structured representations. It brings convenient and efficient solutions to a great number of tasks, such as node classification \cite{ribeiro2017struc2vec}, link prediction \cite{jiao2021temporal}, and knowledge reasoning \cite{zhang2019iteratively}.

\begin{figure}[!t]
\centering

\subfigure[Homogeneous pinwheel]{
\includegraphics[width=0.4\linewidth]{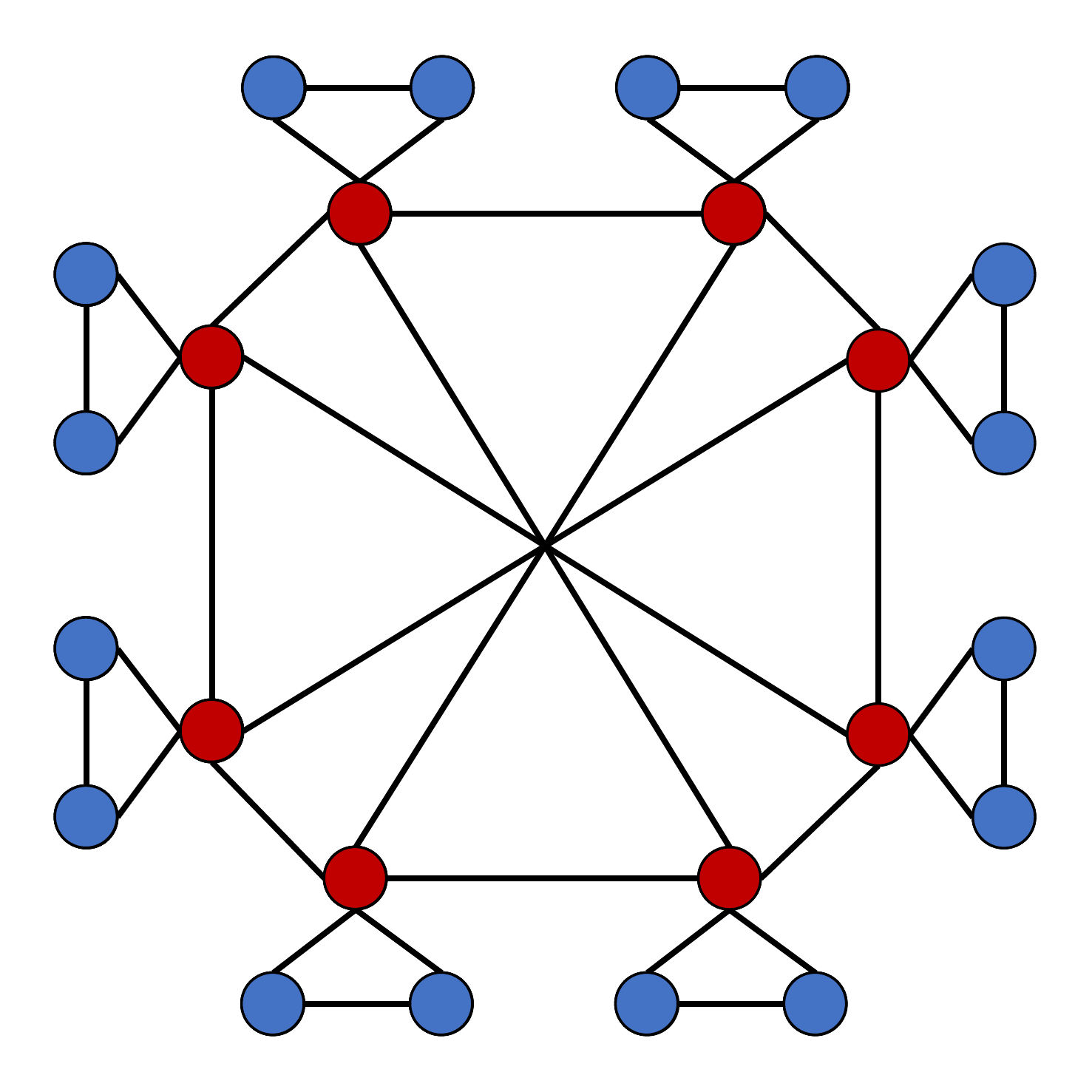}
}
\subfigure[Heterogeneous pinwheel]{
\includegraphics[width=0.4\linewidth]{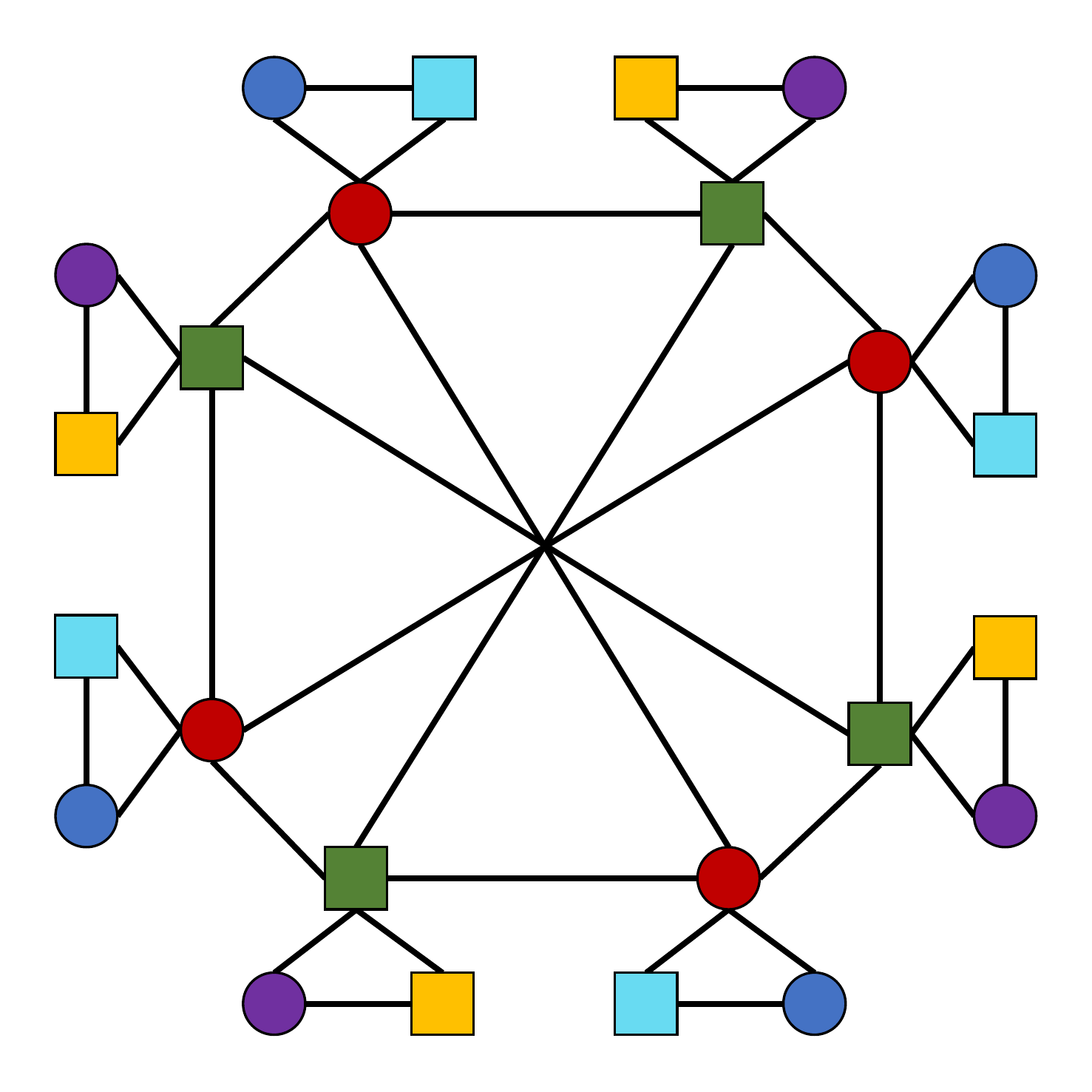}
}

\caption{The synthetic pinwheel networks. Node shapes denote their types and node colors denote their structural roles.}
\label{fig.pinwheel}
\end{figure}

The outbreak of NE research dates back to DeepWalk \cite{perozzi2014deepwalk} that represents nodes in homogeneous networks. Hitherto, on homogeneous networks, almost all the methods have been devised to individually or simultaneously capture two complementary properties \cite{rossi2020proximity}: proximity and structural similarity. The methods on the former, e.g., DeepWalk, aim to preserve the closeness among nodes into embeddings, while the methods capturing the latter, e.g., struc2vec \cite{ribeiro2017struc2vec}, try to make embeddings discriminative on different structural patterns (or roles) \cite{jiao2021survey}. For example, the red nodes and blue nodes having different structures in Fig.~\ref{fig.pinwheel}(a) are in two structural roles. 

In recent years, structural role-based NE has attracted increasing attention, as it can be of great help in learning the function and behavior of nodes \cite{rossi2014role}.
In essence, these methods usually rely on or imitate methods of structural feature extraction \cite{guo2020role,rossi2020structural} and subgraph isomorphism test \cite{ma2019riwalk,nikolentzos2019learning}. It's worth noting that a series of them are developed based on \textbf{anonymous walks (AWs)} \cite{jin2020gralsp,long2020graph,long2021theoretically} to generate structural embeddings or enhance the hot field of Graph Neural Networks (GNNs). Because it's theoretically proved that neighborhood structures of a node can be reconstructed with AWs starting from it \cite{micali2016reconstructing}.


In the real world, heterogeneous networks are more common than homogeneous networks. Because the corresponding nodes of entities in real world usually have multiple types. The heterogeneous structures are much more complex, because the status can be amazingly diverse when different connection patterns meet various node types\footnote{For simplicity, we borrow the electronics term "heterostructures" below to refer to the diverse heterogeneous structures.}. We give an example via Fig. \ref{fig.pinwheel} to demonstrate this fact. The heterogeneous pinwheel has the same connecting patterns as the homogeneous pinwheel. However, with just one more node type, the number of node roles in the former network is three times that of the latter. Therefore, learning representations on heterostructures is a more comprehensive issue, as representing homogeneous structures is only a special case of it.

Nevertheless, little attention has been paid on heterogeneous structural NE. Although NE on heterogeneous networks is also thriving \cite{yang2020heterogeneous}, almost all of existing heterogeneous NE methods can be considered as extensions of those proximity-based homogeneous works in the view of topology. For example, metapath2vec \cite{dong2017metapath2vec} and HIN2vec \cite{fu2017hin2vec} extend DeepWalk and explore meta-paths with biased random walks. R-GCN \cite{schlichtkrull2018modeling}, HetSAGNN \cite{hong2020attention} and HGT \cite{hu2020heterogeneous} are aware of relation types with GNN architectures. Thus, these methods flounder on learning heterostructures. To our best knowledge, node2bits \cite{jin2019node2bits} is the only existing method that tries to fuse both node types and structural features into embeddings. However, it does not consider the combination of node types and underlying structures as a whole (i.e., heterostructure). And its random walk-based feature aggregation and hashing style cannot learn delicate underlying structures.

To fill the big void of NE on heterostructures, we present a terrific amount of work in this paper. The key point is to discriminate the structures mixed with various node types. To this end, we firstly propose a novel technology, \textbf{heterogeneous anonymous walk (HAW)}, where AW is a special case of it on homogeneous networks. And we prove that one can reconstruct heterogeneous neighborhood heterostructures when the distribution of HAWs is known. However, the number of possible HAWs of a given length is usually excessive and in result estimating the exact distribution is practically impossible. To avert this problem, we provide \textbf{heterogeneous anonymous walk embedding (HAWE)} that samples HAWs to capture heterostructures in a data-dirven manner and learns node embeddings by predicting HAWs starting from each node. Additionally, we design coarse HAW (CHAW) for more applicability, and the coarse HAWE (CHAWE) that employs the metioned embedding mechanism on CHAWs. Finally, we conduct sufficient and intuitive experiments on synthetic and real-world networks\footnote{Both the code of our methods and used data in this paper can be found at github.com/naihemeng/HAWE.} that we process for building the first benchmark on heterostructure learning. Compared with both classic homogeneous and heterogeneous NE methods, our HAWE and its variant perform outstandinfly on representing heterostructures. In summary, our contributions are listed as follows:

\begin{itemize}
    \item As far as we know, we are the first to directly study NE that captures heterostructures.
    \item We propose the novel heterogeneous anonymous walk which has theoretical guarantee on reconstructing heterogeneous neighborhood structures, and its more practical variant coarse HAW.
    \item We provide an effective heterogeneous structural NE method HAWE. It generates embeddings by predicting HAWs starting from each node. We also give its variant CHAWE which applies the same mechanism on CHAWs.
    \item We give the first benchmark on heterostructure learning. The sufficient and intuitive experiments shows that our methods achieves excellent performance on distinguishing heterostructures and can be applied on large-scale networks. 
\end{itemize}

\section{Preliminaries}
\label{sec:preliminaries}
\begin{figure*}[!t]
\centering
\includegraphics[width=\linewidth]{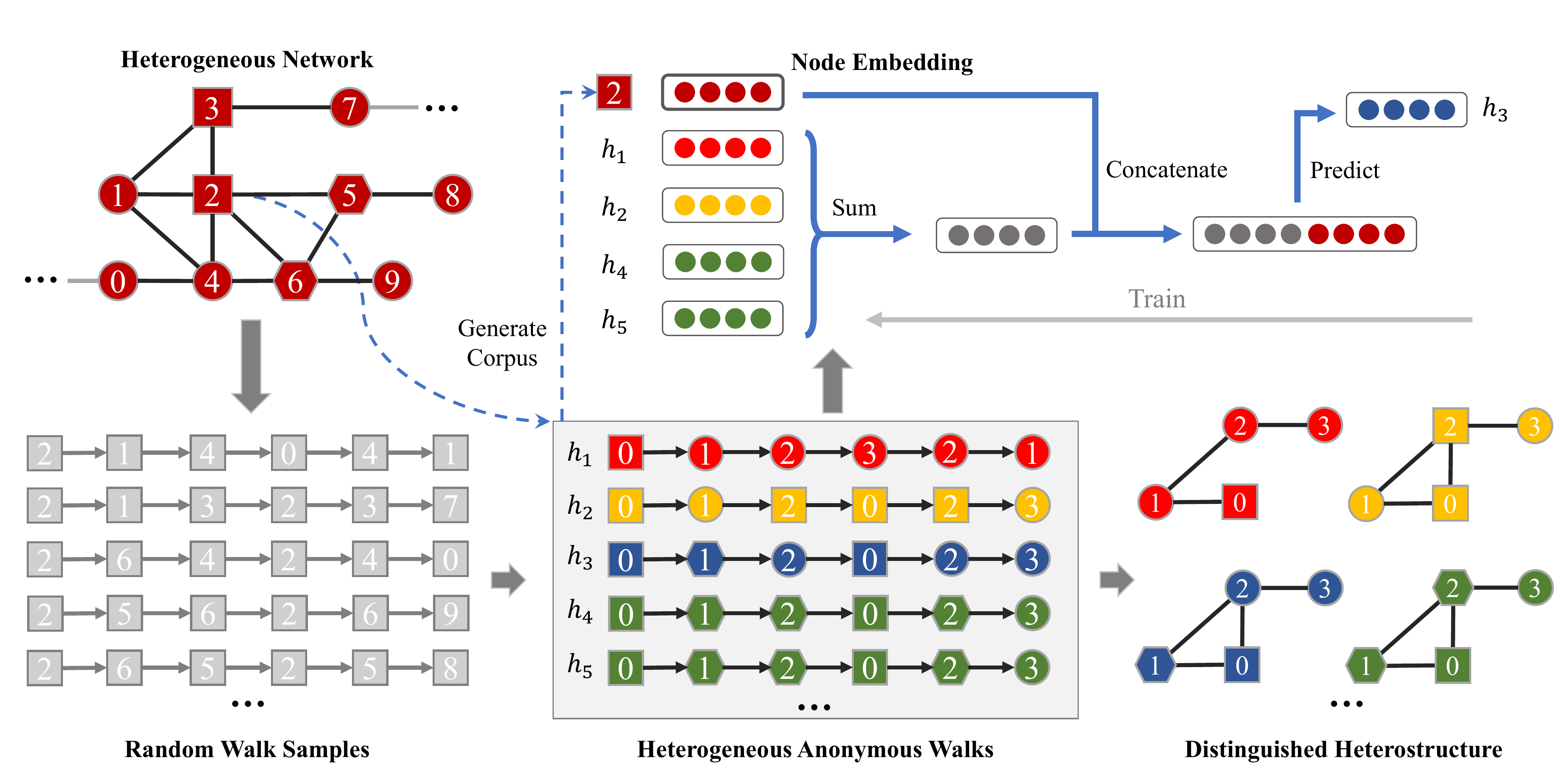}
\caption{The examples of heterogeneous anonymous walk (HAW) and the overview of proposed heterogeneous anonymous walk embedding (HAWE). The node shapes denote node types. The walks in the same color ($h_4$ and $h_5$) capturing the same heterostructure.}
\label{fig.model}
\end{figure*}

\subsection{Problem Definition}
\begin{definition}[\textbf{Homogeneous Network}] A homogeneous network is represented as a graph $\mathcal{G} = (\mathcal{V},\mathcal{E})$, where $\mathcal{V}$ is the set of nodes and $\mathcal{E} \subseteq \mathcal{V} \times \mathcal{V}$ is the set of edges. 
\end{definition}

\begin{definition}[\textbf{Heterogeneous Network}] A heterogeneous network is defined as $\mathcal{H} = (\mathcal{V}, \mathcal{E}, \Phi, \Psi)$. Here nodes $\mathcal{V}$ and edges $\mathcal{E}$ construct the topology. $\Phi$ maps each node $v \in \mathcal{V}$ to a node type $\Phi(v) \in \mathcal{T}$ where $\mathcal{T}$ is the node type set. And $\Psi$ maps each edge $e\in\mathcal{E}$ to an edge type $\Phi(e) \in \mathcal{T}_{\mathcal{E}}$. 

\end{definition}

Note that the type of an edge is determined by the node type of its two ends in most heterogeneous networks. We omit edge types in this paper.

\begin{definition}[\textbf{Network Embedding}] Given a homogeneous/heterogeneous network $\mathcal{G}$/$\mathcal{H}$, network embedding (NE) aims to map $\mathcal{V}$ to low-dimensional vector representations $\{ \mathbf{z}_v \in \mathbb{R}^d | v \in \mathcal{V} \}$ where the embedding dimension $d$ is usually much smaller than $|\mathcal{V}|$. 
\end{definition}

We don't learn different mappings for different types of nodes in this paper, as we consider node types are an important part of the heterostructures. 

\subsection{Anonymous Walks}
\begin{definition}[\textbf{Random Walk}]
A random walk is a sequence of nodes $w = (v_0, v_1, ..., v_l)$ where each latter node $v_{i+1}$ is sampled randomly from the neighbor set of the previous one $\mathcal{N}(v_i)$, $ 0 < i \le l$. The length of $w$ is $l$. 
\end{definition} 

Given a network, generating random walks is an effective and efficient way to sample subsets of nodes and edges. While learning the structural patterns formed by the sampled nodes and edges, the node Id information is redundant. Thus, an anonymization process on random walks is usually applied. 

\begin{definition}[\textbf{Anonymous Walk}]
Given a random walk $w = (v_0, v_1, ..., v_l)$, the corresponding anonymous walk (AW) of it is the $l$-length sequence of integers $a = (f(v_0), f(v_1), ..., f(v_l)) $. $f$ is the position function over $w$ defined as $f(v_i,w) = |\{v_0, ..., v_{i'}\}|$ where $\{v_0, ..., v_{i'}\}$ is a subsequence of $w$ starting from $v_0$ and $i'$ is the smallest integer $j$ satisfying $v_j = v_i$.  For simplicity, we omit $w$ later when denoting the position of a node in it by $f$.
\end{definition}

See the lower-left part of Fig. \ref{fig.model} for some AW examples where the node types are ignored. After anonymization, the shown $5$ different random walks are mapped to $2$ AWs $(0,1,2,3,2,1)$ and $(0,1,2,0,2,3)$. Meanwhile, the $2$ AWs can be regarded as $2$ kinds of subgraphs: $4$-path and tailed-triangle, respectively. Therefore, with anonymous walks, we can learn higher-order structures similar to graphlets which are considered as the building blocks of networks and node functions \cite{milo2002network}. 

\begin{theorem}
\label{theorem:aw_num}
The number of all possible AWs of the given length $l$ is $B_l$. $B_l$ is the $l$-th Bell number that can be computed recurrently: $B_l = \sum_{k=0}^{l-1} \tbinom{l-1}{k} B_k$, where $B_k$ starts from $B_0 = B_1 = 1$. 
\end{theorem}

\begin{proof}[Proof]
  Consider anonymizing the occurred edges instead of nodes in a given AW so that each AW can be translated to a unique new sequence. For example, the AW $(0,1,2,0,1)$ now can be represented as $(0, 1, 2, 0)$. Then, the counting possible AWs is equivalent to the problem of counting rhyme schemes \cite{gardner1978bells}, by which the theorem can be proved.
\end{proof}

\begin{theorem} 
\label{theorem:rcst_g}
\cite{micali2016reconstructing} Let $\mathcal{G}_r(v)$ be a homogeneous subgraph of which all nodes are within $r$ distance from node $v\in \mathcal{V}$. Then one can reconstruct $\mathcal{G}_r(v)$ with $P_l$, where $P_l$ is the distribution of $l$-length anonymous walks starting from $v$ with $l = 2(m+1)$, and $m$ is the number of edges in  $\mathcal{G}_r(v)$.
\end{theorem}

Theorem~\ref{theorem:rcst_g} guarantees that structures can be captured with AWs. However, according to Theorem~\ref{theorem:aw_num}, estimating the exact distribution of AWs is hard, even though the used AWs are not very long. Thus, sampling strategies are always used on sizeable scale networks in previous AW-based NE works \cite{jin2020gralsp,long2020graph,long2021theoretically}.

\section{Method}
\label{sec:method}

\subsection{Heterogeneous Anonymous Walks}
As stated, we are interested in learning the heterostructure, a mixture of different structural patterns and node types. Inspired by AWs having the excellent ability to capture structures, we consider proposing a technique that can capture underlying structures as AWs do and identify the node types at the same time. Thus, we give the following definition of heterogeneous anonymous walk.

\begin{definition}[\textbf{Heterogeneous Anonymous Walk}]
Given a random walk $w = (v_0, v_1, ..., v_l)$ occurring in a heterogeneous network $\mathcal{H} = (\mathcal{V}, \mathcal{E}, \Phi)$, its corresponding heterogeneous anonymous walk (HAW) is the $l$-length sequence of 2-tuples $h = (g(v_0), g(v_1), ..., g(v_l))$, where the tuple $g(v_i) = (f(v_i), \Phi(v_i))$.
\end{definition}

As shown in the bottom part of Fig. \ref{fig.model}, the $4$ random walks that mapped to the same AW $(0,1,2,0,2,3)$ now can be distinguished as $3$ unique HAWs. Each of these HAWs can be used to construct a heterogeneous subgraph. Although the $3$ constructed subgraphs share the same underlying structures, i.e., tailed triangle, they are different because of the different distributions of node types. We can still consider the constructed subgraphs as a new kind of graphlets. Empirically, our HAWs are  equivalent to position-aware typed graphlets \cite{rossi2020heterogeneous} which are aware of not only the type but also the positions of each node.

More theoretically, we show that HAWs can be used to reconstruct the neighborhood heterostructures of a given node by the following theorem.

\begin{theorem} 
\label{theorem:rcst_h}
Let $\mathcal{H}_r(v)$ be a heterogeneous subgraph of which all nodes are within $r$ distance from node $v\in \mathcal{V}$. One can reconstruct $\mathcal{H}_r(v)$ with $Q_l$, where $Q_l$ is the distribution of $l$-length heterogeneous anonymous walks starting from $v$ with $l = 2(m+1)$, and $m$ is the number of edges in  $\mathcal{H}_r(v)$. 
\end{theorem}

\begin{proof}[Proof]
Here we just provide a non-constructive proof since our aim is not to design a practical reconstruction algorithm. 

We call a HAW $h$ economical if any ordered pairs $((s,\Phi(f^{-1}(s))), (t,\Phi(f^{-1}(t))))$ occurs at most once in it. Define $\mathcal{S}=\bigcup_{i=1}^{l} supp(Q_i)$. Let $H(h) = (V(h), E(h), \Phi')$ be the heterogeneous network reconstructed from $h = (g(v_0), g(v_1), ..., g(v_l))$, where $V(h) = \{f(v_i) | 0 \le i \le l \}$, $E(h) = \{ (f(v_i),f(v_{i+1})) | 0 \le i < l \}$, and $\Phi'$ satisfies $\Phi'(s) = \Phi(f^{-1}(s))$. Then we have $h\in \mathcal{S} \iff H(h) $ is isomorphic to a heterogeneous subgraph of $\mathcal{H}$ and the start node $v_0$ of $h$ is mapped to $f(v_0)=0$ in $H(h)$. Thus, we can enumerates over $\mathcal{S}$ to find the longest $h^*\in \mathcal{S}$ such that both $h^*$ is economical and $H(h^*)$ is of radius $r$ from its central node $0$.  It's intuitive that $H(h^*)$ is isomorphic to  $\mathcal{H}_r(v_0)$ which is a subgraph of $\mathcal{H}$, too. The $h^*\in \mathcal{S}$ must exist as any the longest economical $h$ of length $2(m+1)$ covers at least $m+1$ edges of its $H(h)$.
\end{proof}

Though HAWs can be used 
to capture the heterostructure of each node in light of Theorem \ref{theorem:rcst_h}, the ideal situation for achieving this is that we know the exact distribution of HAWs. Therefore, we wonder how difficult it is to estimate the exact HAW distribution.

\begin{theorem} 
\label{theorem:haw_num}
The number of all possible HAWs of the given length $l$ is $|\mathcal{T}|^l B_l$ where there are $|\mathcal{T}|$ node types in the given heterogeneous network.
\end{theorem}

According to Theorem \ref{theorem:haw_num}, the number of all possible HAWs grows exponentially w.r.t. walk length and much more rapider than that of AWs. Although $|\mathcal{T}|^l B_l$ is just the upper bound for a specific heterogeneous network, computing HAW distribution is almost impossible, since the distribution of node types is always complex. Thus, we choose to use a sampling strategy to more efficiently take advantage of HAWs.

\subsection{Heterogeneous Anonymous Walk Embedding}
In this part, we consider how to preserve the heterostructure information captured by each HAW sample of a node into the corresponding embedding. 
To solve the problem, inspired by AWE \cite{ivanov2018anonymous}, a method learning embeddings representing the whole graphs, we imitate the way the language model PV-DM \cite{le2014distributed} generates paragraph representations. 

Specifically, we treat the neighborhood of each node as a text paragraph and the sampled HAWs starting from the same node as the context words that occur together in the corresponding paragraph. The neighborhood heterostructures of a node usually have a theme. For example, an active user in a question-and-answer network behaves like a star-center node because of his/her large amounts of activities, i.e. proposing questions and answers, while an expert user may have much less activities but propose much more answers than questions. The heterostructure theme can be expressed by the context words, i.e., sampled HAWs and makes it possible to predict an word in the context when knowing some other context words. Thus, we can construct an embedding model by learning this predictability.

Based on the above idea, we propose heterogeneous anonymous walk embedding (HAWE), a method to learn node representations on heterostructures. Its overview illustration is shown in Fig. \ref{fig.model}. 

Like learning language models, we need to generate a heterostructure corpus first. For each paragraph node $v \in \mathcal{V}$ in a given heterogeneous network $\mathcal{H}$, we sample a sequence of $T$ HAWs $C^v = (h_1^v, h_2^v,..., h_T^v)$ starting from it as its context words. Let $\mathcal{L}$ be the HAW lexicon over the corpus $\{C^v \}_{v \in \mathcal{V}}$. The size of $\mathcal{L}$ is always smaller than that of the total corpus, since there are many common HAWs over all the contexts. And for different nodes, it is their common HAWs that reflect the neighborhood heterosturcture similarities between them.

On the corpus $\{C^v \}_{v \in \mathcal{V}}$, HAWE learns a set of node embeddings $\{\mathbf{z}_v \in \mathbb{R}^d \}_{v\in\mathcal{V}}$ as well as a set of HAW embeddings $\{\mathbf{w}_h \in \mathbb{R}^d \}_{h \in \mathcal{L}}$, where $d$ is embedding dimension of both node embeddings and HAW embeddings. For simplicity, we denote the HAW embeddings corresponding to the HAWs in sequence $C^v = (h_1^v, h_2^v,..., h_T^v)$ as $ \mathbf{w}_1^v, \mathbf{w}_2^v ..., \mathbf{w}_T^v$. HAWE establishes the relations among the paragraphs (i.e., nodes) and words (i.e., HAWs) by predicting words based on the paragraphs they belong to and their contexts sampled via a sliding window of length $\Delta$. More formally, we train HAWE by maximizing the average log probabilities for all the words occurring in the corpus $\{C^v \}_{v \in \mathcal{V}}$ as follows:
\begin{equation}
    \frac{1}{|\mathcal{V}|} \frac{1}{T} \sum_{v\in\mathcal{V}} \sum_{t=\Delta}^{T-\Delta}\log p(\mathbf{w}_t^v | \mathbf{w}_{t-\Delta}^v,...,\mathbf{w}_{t+\Delta}^v, \mathbf{z}_v),
    \label{eq:prediction}
\end{equation}
where each probability is computed via softmax function:
\begin{equation}
    p(\mathbf{w}_t^v | \mathbf{w}_{t-\Delta}^v,...,\mathbf{w}_{t+\Delta}^v, \mathbf{z}_v) = \frac{e^{y(\mathbf{w}_t^v)}}{\sum_{h\in \mathcal{L}}e^{y(\mathbf{w}_h)}},
    \label{eq:probability}
\end{equation}
The un-normalized prediction probability $y(\mathbf{w}_{t}^v)$ for word $h_t^v$ is computed as follows:
\begin{align}
    \widehat{\mathbf{w}}_t^v & = \mathrm{Sum}(\mathbf{w}_{t-\Delta}^v,...,\mathbf{w}_{t+\Delta}^v), \label{eq:sum}\\
    y(\mathbf{w}_{t}^v) &= b + \mathbf{u}^\top[\widehat{\mathbf{w}}_t^v, \mathbf{z}_v],
\end{align}
where $b \in \mathbb{R}$ and $\mathbf{u} \in \mathbb{R}^{2d}$ are learnable parameters. Here we sum over the context embeddings together to preserve as much heterostructure information as possible. Because we consider Eq. (\ref{eq:sum}) as the pooling mechanism that Graph Neural Networks need to apply after massage-passing. And it is proved sum-pooling is more effective that other simple poolings such as mean- and max-pooling \cite{XuHLJ19, guo2020role}. 

In light of above designs, the node embedding $\mathbf{z}_v$ is shared only across the contexts in which all the HAWs start from node $v$ and has nothing to do with other contexts. Therefore, the different heterostructures can be differentiated by the node embeddings $\{\mathbf{z}_v \}_{v\in\mathcal{V}}$ trained via maximizing Eq. (\ref{eq:prediction}). However, computing the denominator part of Eq. (\ref{eq:prediction}) needs very high cost in practice. To mitigate this problem, we use hierarchical softmax\cite{mikolov2013distributed} to speed up the computation by replacing the multi-class classification task with multi layers of binary classification tasks.

\subsection{Coarse HAW and HAWE}
In ideal situation, the propsed HAWE can distinguish the heterostructure captured by each unique HAW. However, the design of HAW isn't the most practical in many cases. On the one hand, the node embeddings capture similarities based on the common HAWs occurring in different contexts while ignoring the similarities between the heteorstructures captured by different HAWs. On the other hand, on large-scale networks, the biggest HAW sample size for acceptable
efficiency is still much smaller than the sample size required for a full understanding of the entire heterogeneous structure (Theorem \ref{theorem:haw_num}). To deal with these problems, we propose a more practical variant of HAW. 

\begin{table*}[!t]
\caption{Statistics of real-world heterogeneous networks. Abbreviations of node types: A for airport (in Air-traffic)/answer (in Stack Exchange networks), C for country, Q for question, U for user. Abbreviations of Stack Exchange topics: Anime for anime \& manga, CG for computer graphics, Chem for chemistry, CSE for computer science educators, Engr for engineering, FIT for physical fitness, HWR for hardware recommendations, IOT for Internet of Things, Latin for Latin language, Movie for movies \& TV.}
\centering
\begin{tabular}{cccc}
\toprule
\textbf{Dataset} & \textbf{$\#$nodes} & \textbf{$\#$edges} & \textbf{$\#$classes} \\
\midrule
\textbf{Air-traffic} & A: 3,373, C: 226 $|$ total: 3,599 & A-A: 19,150, A-C: 3,373 $|$ total: 22,523  & A: 2 \\
\textbf{SE-Anime} & A: 1,398, Q: 696, U: 236 $|$ total: 2,330  & A-Q: 1,398, A-U: 1,398, Q-U: 183 $|$ total: 2,979 & A: 3, U: 3 \\
\textbf{SE-Beer} & A: 2,343, Q: 1,005, U: 1,161 $|$ total: 4,509 & A-Q: 2,343, A-U: 2,343, Q-U: 589 $|$ total: 5,275 & A: 4, U: 3 \\
\textbf{SE-CG} & A: 3,186, Q: 2,326, U: 1,902 $|$ total: 7,414 & A-Q: 3,186, A-U: 3,186, Q-U: 1,652 $|$ total: 8,024  & A: 3, U: 4 \\
\textbf{SE-Chem} & A: 1,651, Q: 802, U: 402 $|$ total: 2,855 & A-Q: 1,651, A-U: 1,651, Q-U: 318 $|$ total: 3,620  & A: 3, U: 4 \\
\textbf{SE-CSE} & A: 3,944, Q: 920, U: 1,230 $|$ total: 6,094 & A-Q: 3,944, A-U: 3,944, Q-U: 435 $|$ total: 8,323 & A: 4, U: 3 \\
\textbf{SE-Engr} & A: 14,935, Q: 9,084, U: 8,098 $|$ total: 32,117 & A-Q: 14,935, A-U: 14,935, Q-U: 6,871 $|$ total: 36,741 & A: 4, U: 4 \\
\textbf{SE-FIT} & A: 16,998, Q: 8,309, U: 6,996 $|$ total: 32,303 & A-Q: 16,998, A-U: 16,998, Q-U: 4,575 $|$ total: 38,571 & A: 5, U: 3 \\
\textbf{SE-HWR} & A: 2,854, Q: 1,953, U: 2,723 $|$ total: 7,530 & A-Q: 2,854, A-U: 2,854, Q-U: 2,409 $|$ total: 8,117 & A: 4, U: 3 \\
\textbf{SE-IOT} & A: 2,250, Q: 1,620, U: 1,354 $|$ total: 5,224 & A-Q: 2,250, A-U: 2,250, Q-U: 1,280 $|$ total: 5,780 & A: 3, U: 3 \\
\textbf{SE-Latin} & A: 6,636, Q: 4,087, U: 1,655 $|$ total: 12,378 & A-Q: 6,636, A-U: 6,636, Q-U: 1,339 $|$ total: 14,611 & A: 4, U: 3 \\
\textbf{SE-Movie} & A: 2,015, Q: 1,107, U: 481 $|$ total: 3,603 & A-Q: 2,015, A-U: 2,015, Q-U: 402 $|$ total: 4,432 & A: 3, U: 3 \\

\bottomrule
\end{tabular}

\label{tab.dataset}
\end{table*}

\begin{definition}[\textbf{Coarse Heterogeneous Anonymous Walk}] For a HAW $h= (g(v_0), g(v_1), ..., g(v_l))$ occurring in the given heterogeneous network $\mathcal{H} = (\mathcal{V}, \mathcal{E}, \Phi)$, its corresponding coarse heterogeneous anonymous walk (CHAW) is the 2-tuple $c = (a, \mathrm{OC}(w))$ composed of the $l$-th AW $a = (f(v_0), f(v_1), ..., f(v_l))$ and the ordered node type count $\mathrm{OC}(w) = ((\phi_0, n_0), (\phi_1, n_1),..., (\phi_O, n_O))$. $\phi_i$ is the $i$-th earliest kind of node type occurring in $h$ and $n_i$ is its presence frequency. $O \le |\mathcal{T}l|$ is the number of all kinds of node types in $h$.
\end{definition}

CHAWs count the node types and still remain a little but important position information by the order of the count list. In the view of graphlets, our CHAWs represent more delicate heterostructures than typed graphlets \cite{rossi2020heterogeneous} which only care about node type presence frequency but totally ignore their positions, and less delicate heterostructures than position-aware typed graphlets. 

Intuitively, lots of HAWs capturing similar heterostructures are grouped and represented by a single CHAW so that the number of all possible CHAWs is much smaller than that of all possible HAWs. Thus, we can alleviate the above problems by replace the HAWs in HAWE with corresponding CHAWs and get the variant of embedding model CHAWE.

\subsection{Time Complexity}
\label{sec:time}
Sampling a $L$-length HAW/CHAW costs time of $O(L)$. Thus, the time complexity of obtaining heterogeneous corpus is $O(LT|\mathcal{V}|)$, as there are $T$ HAWs/CHAWs sampled for each node. It takes $O(dT\Delta|\mathcal{V}|\log|\mathcal{L}|)$ for HAWE/CHAWE to learn representations from the heterogeneous corpus. Compared with HAWE, CHAWE usually gets a smaller lexicon $\mathcal{L}$ so that it costs less time to generate embedddings. Considering the sparsity of real-world networks and most heterogeneous network having a simple schema graph with no selfloops (i.e., there is no node having neighbors with the same type), $|\mathcal{L}|$ is usually much smaller than $T|\mathcal{V}|$. Therefore, the overall complexity of HAWE/CHAWE is almost linear to the number of nodes.

\section{Experiments}
\label{sec-experiments}
\begin{figure*}[!t]
\centering

\subfigure[DeepWalk]{
\includegraphics[width=0.18\linewidth]{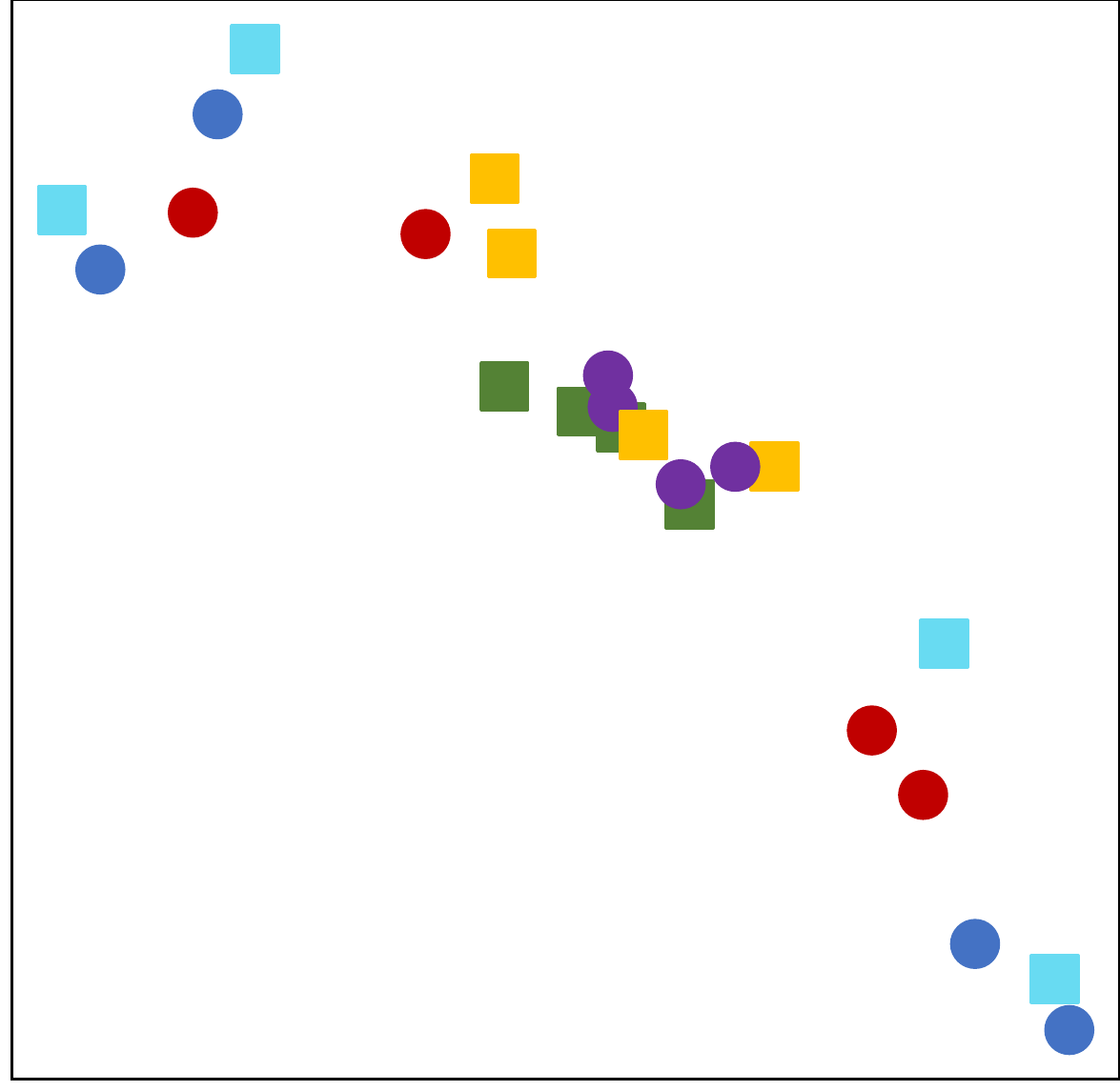}
}
\subfigure[LINE]{
\includegraphics[width=0.18\linewidth]{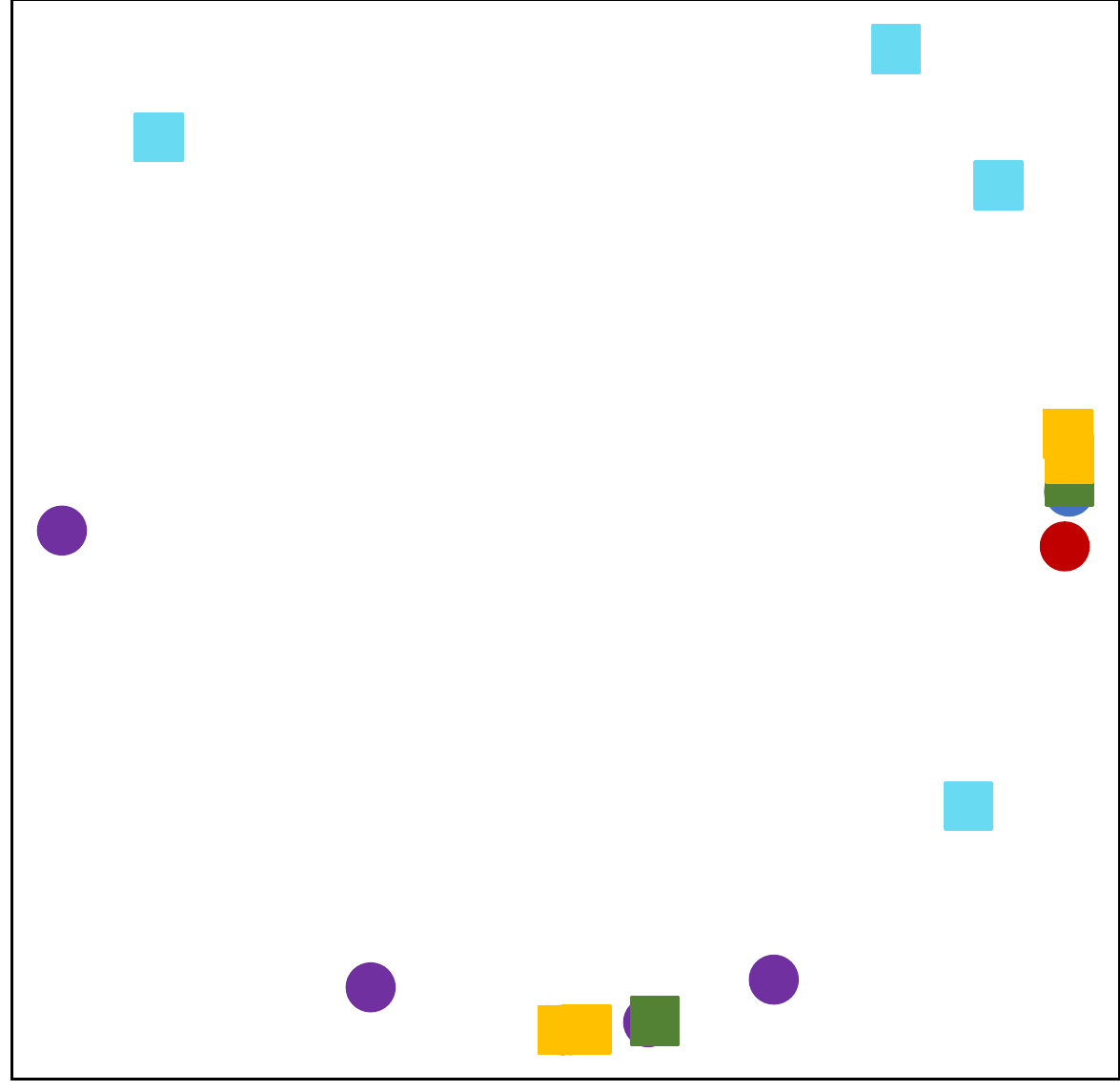}
}
\subfigure[RolX]{
\includegraphics[width=0.18\linewidth]{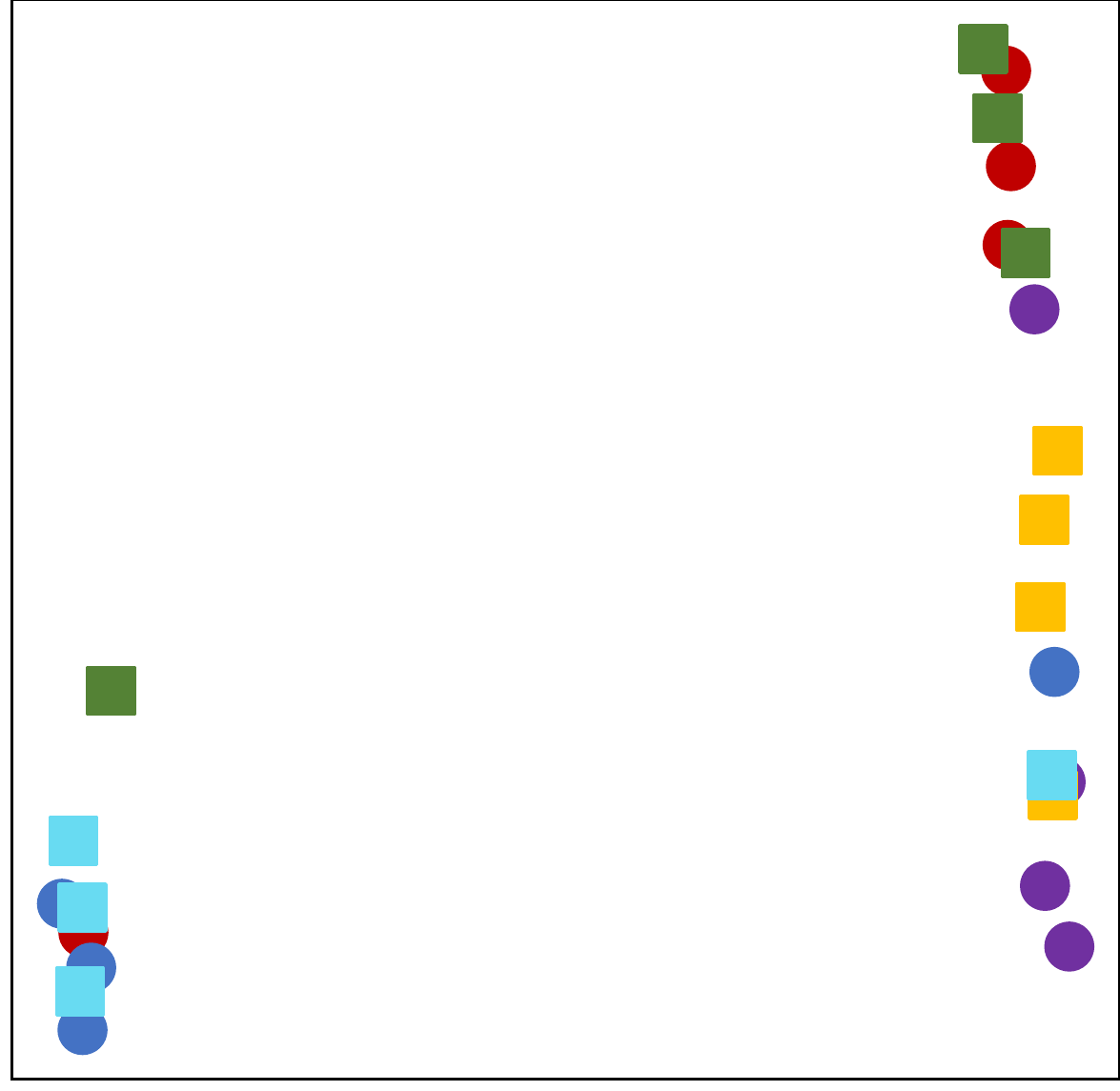}
}
\subfigure[struc2vec]{
\includegraphics[width=0.18\linewidth]{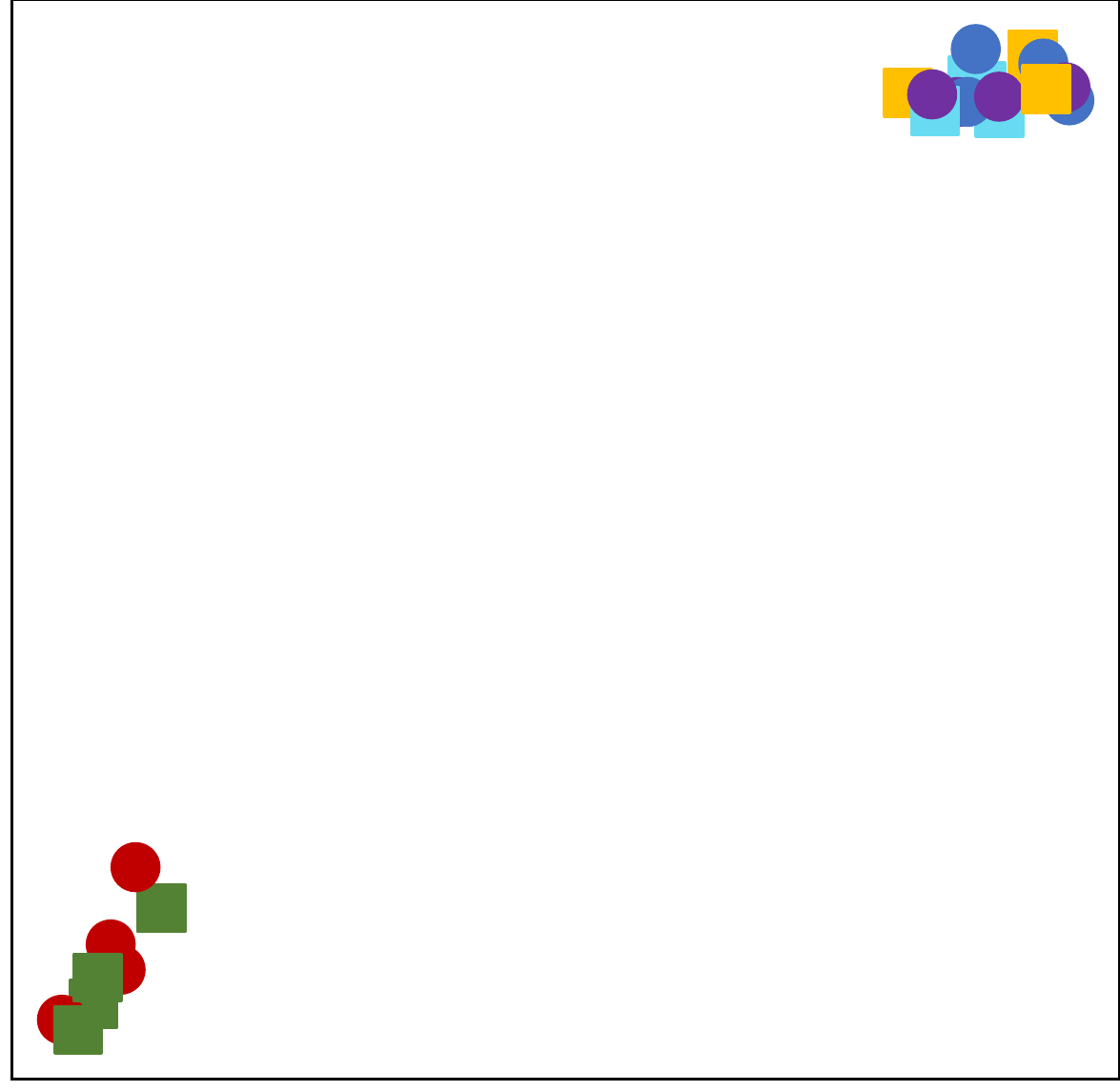}
}
\subfigure[GraphWave + PCA]{
\includegraphics[width=0.18\linewidth]{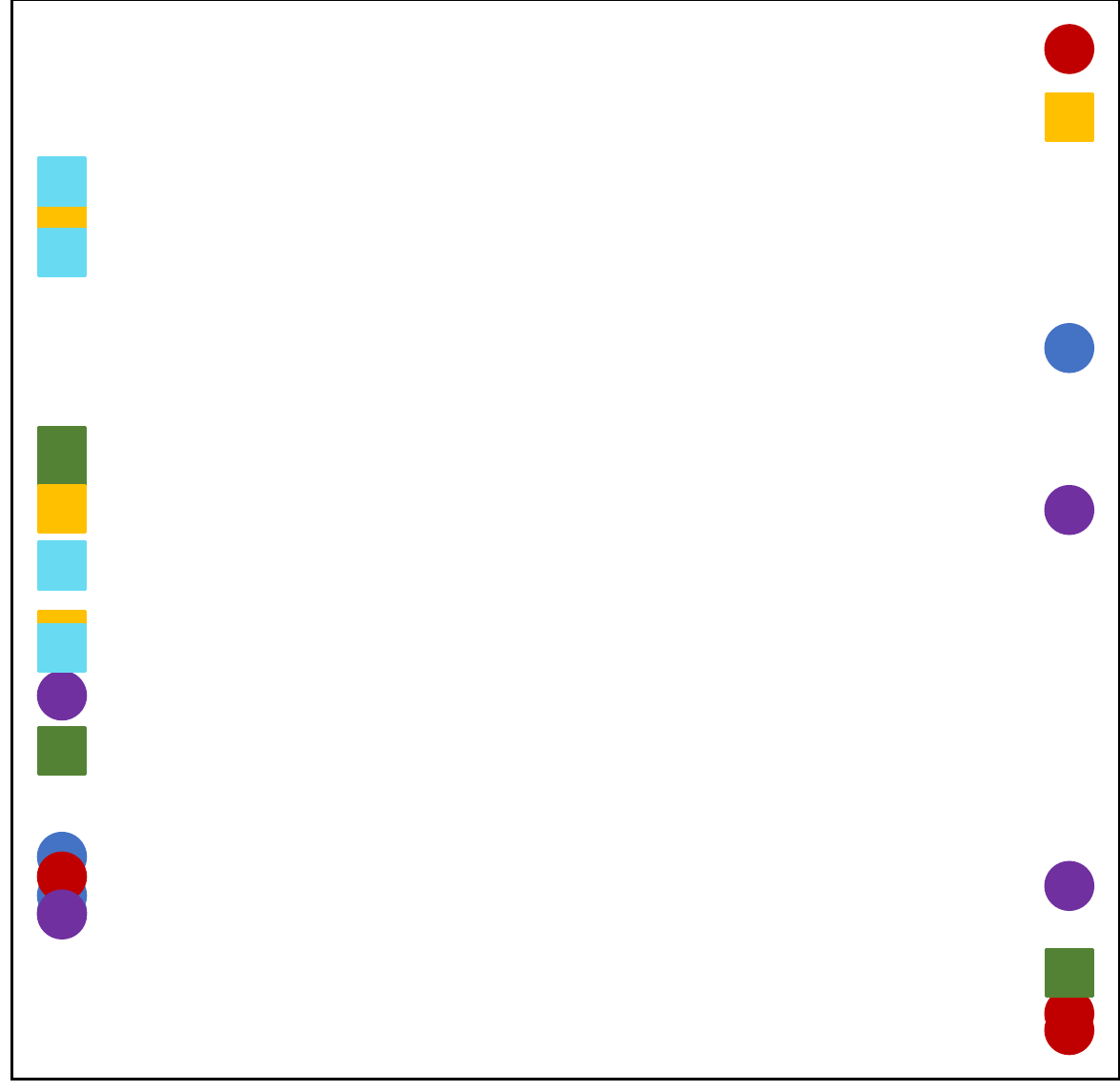}
}\\
\subfigure[role2vec]{
\includegraphics[width=0.18\linewidth]{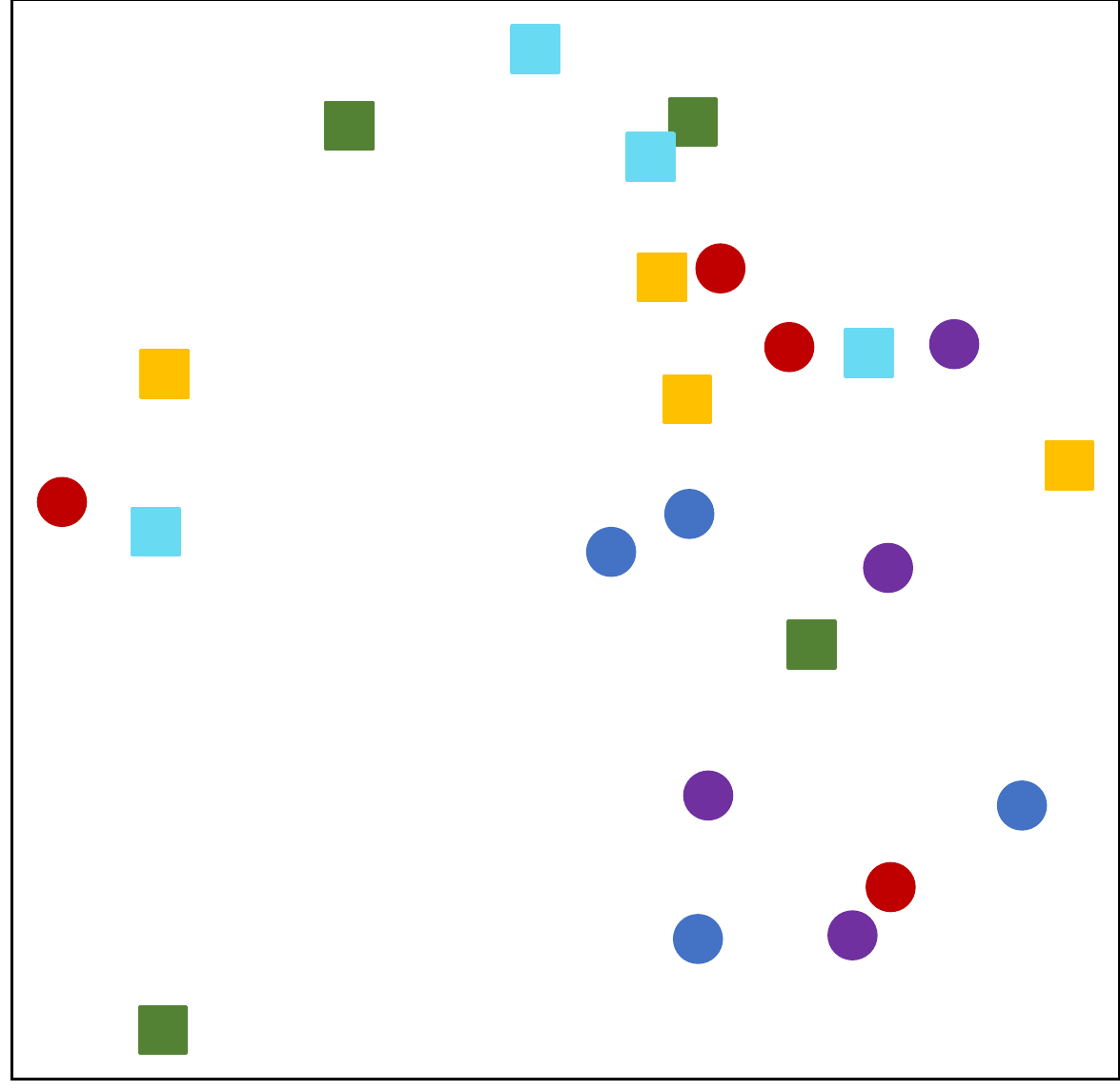}
}
\subfigure[GraphSTONE]{
\includegraphics[width=0.18\linewidth]{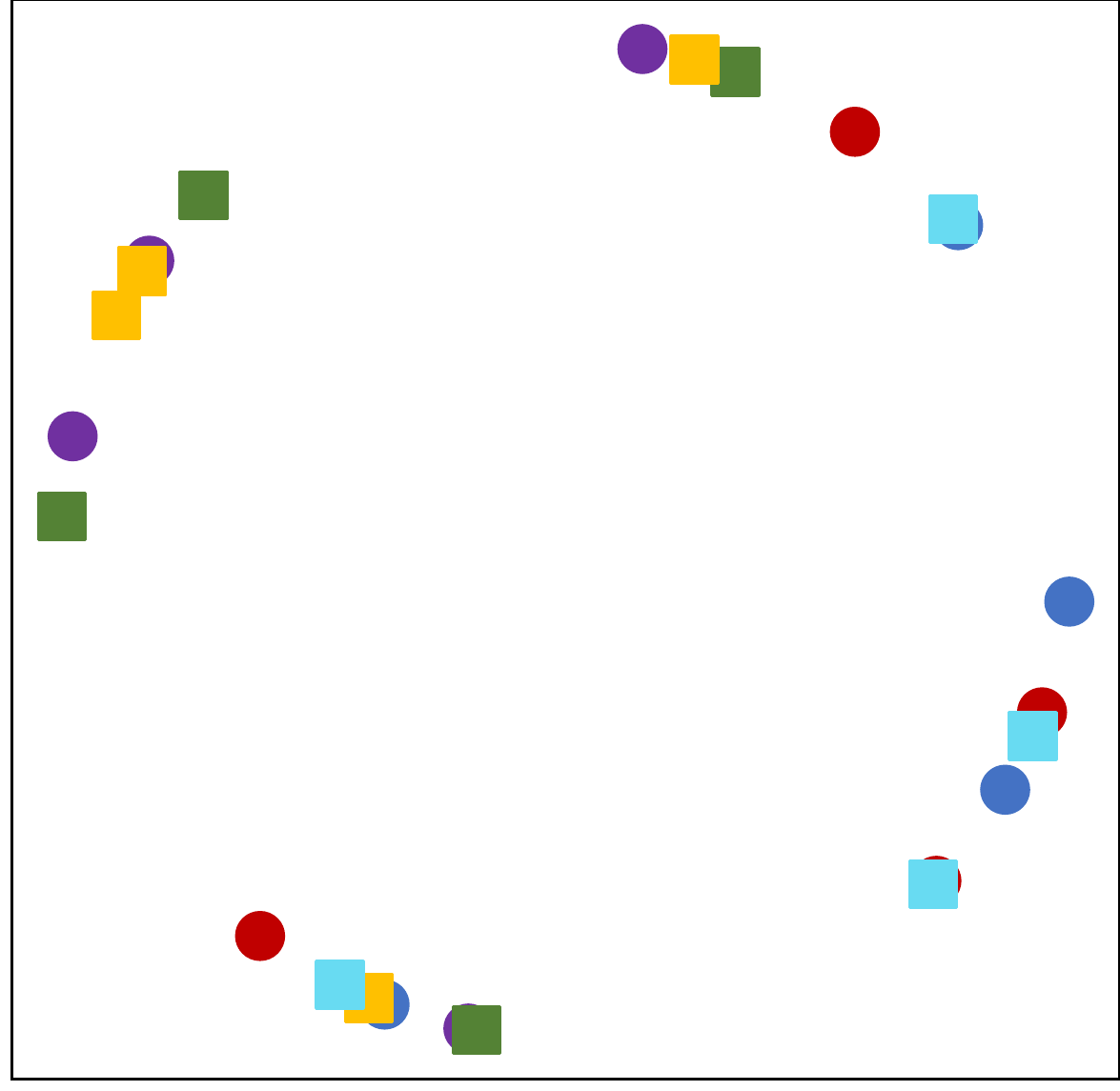}
}
\subfigure[HIN2vec]{
\includegraphics[width=0.18\linewidth]{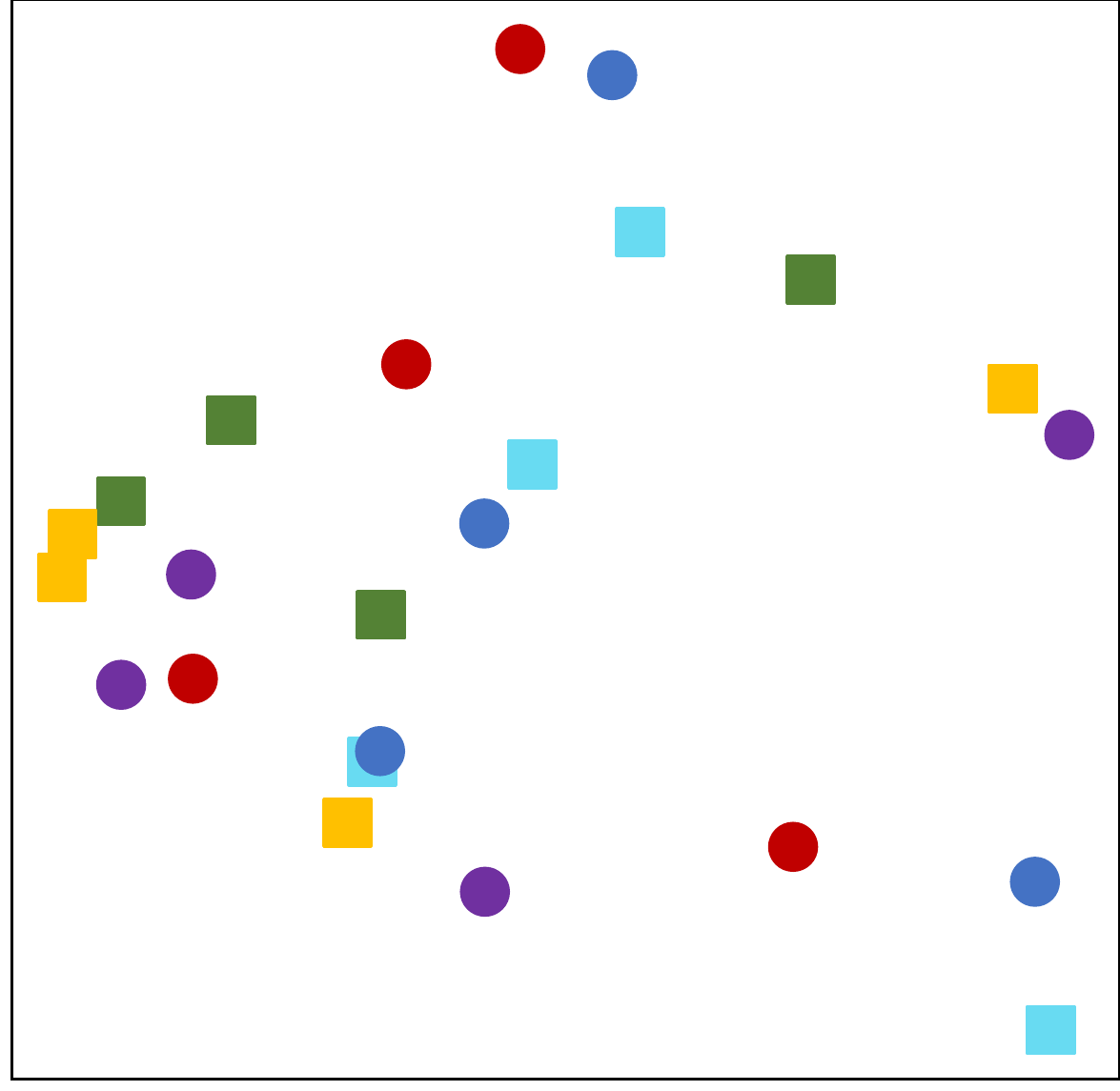}
}
\subfigure[TransE]{
\includegraphics[width=0.18\linewidth]{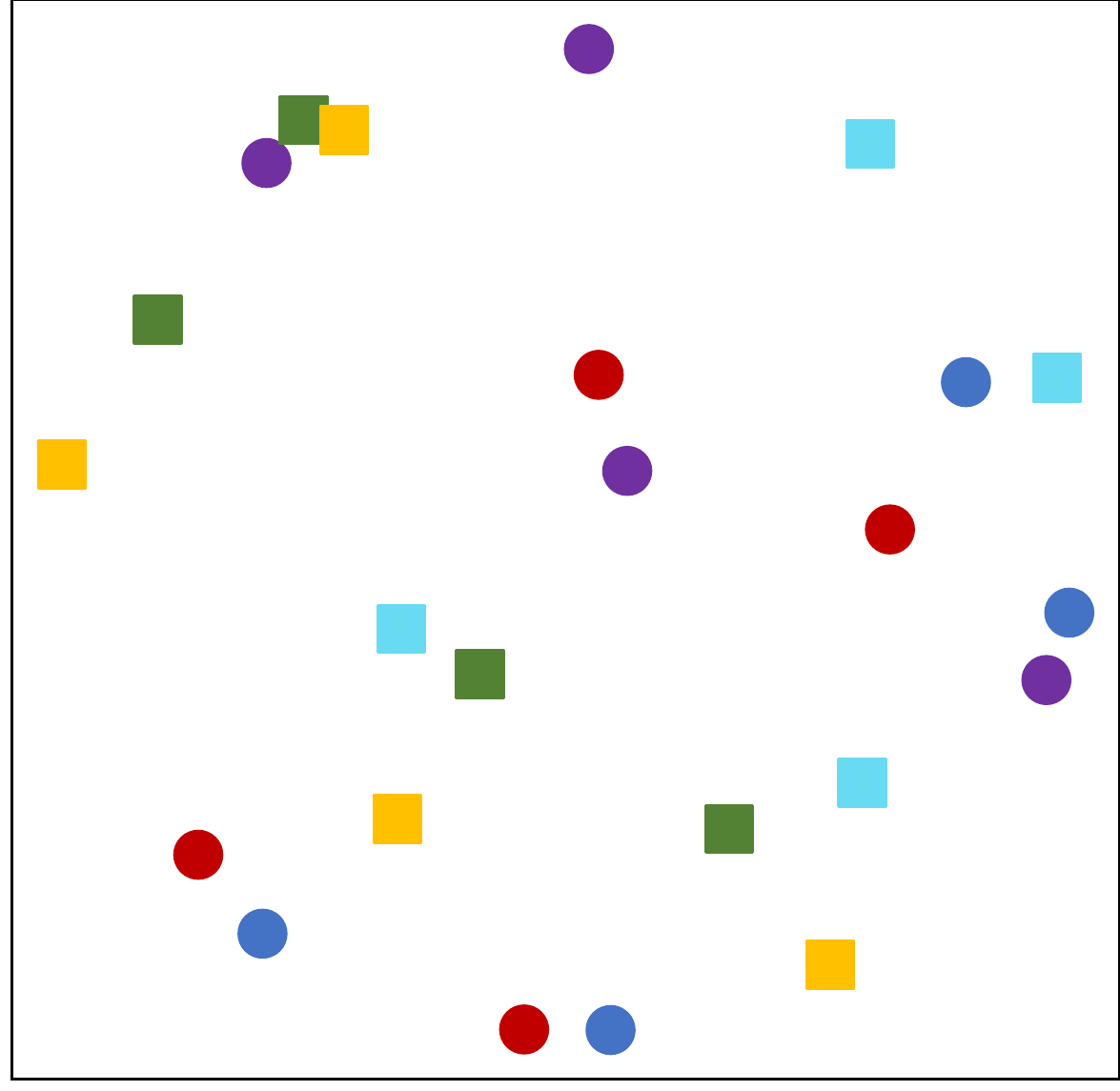}
}
\subfigure[node2bits + PCA]{
\includegraphics[width=0.18\linewidth]{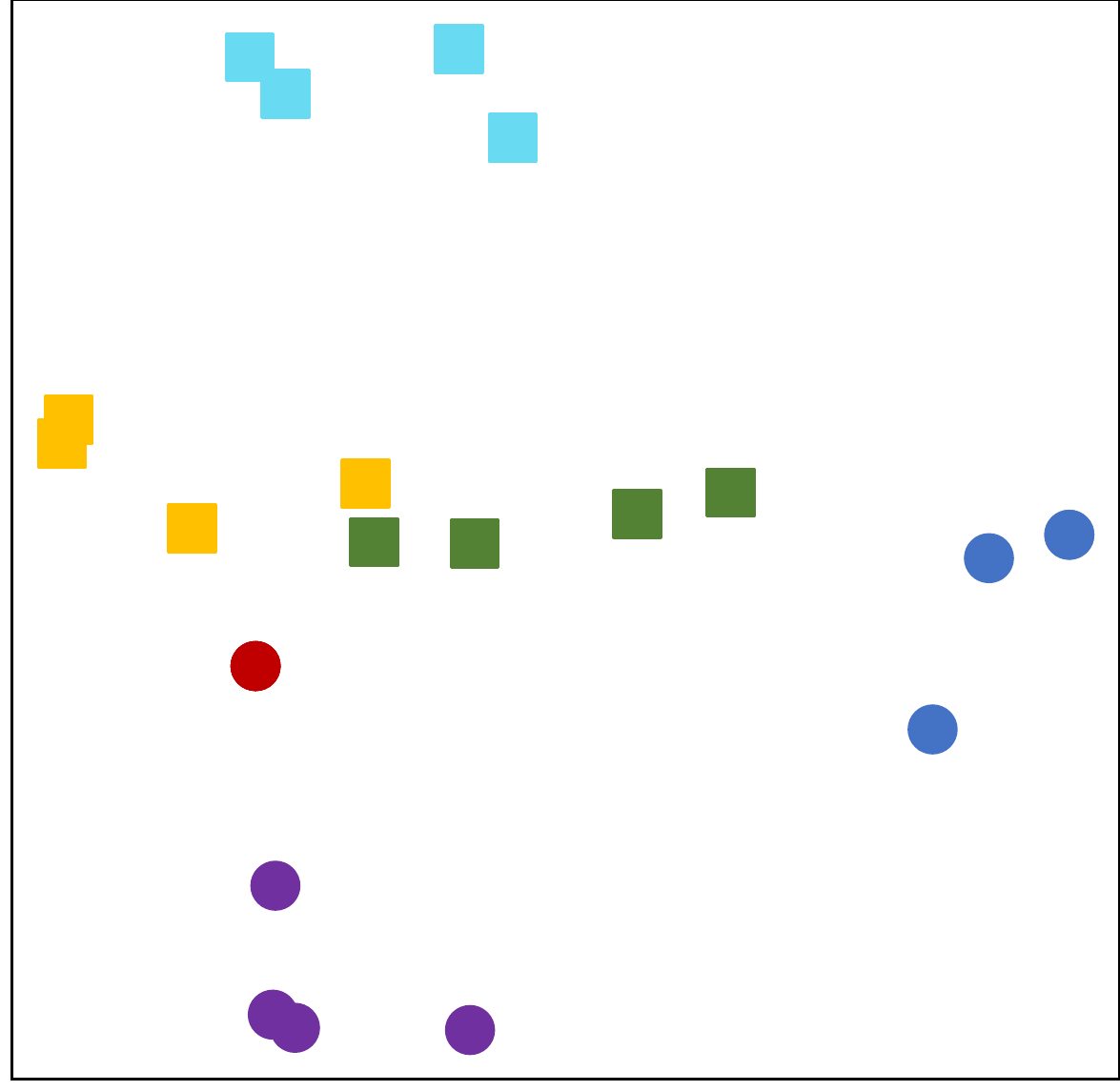}
}\\
\subfigure[R-GCN]{
\includegraphics[width=0.18\linewidth]{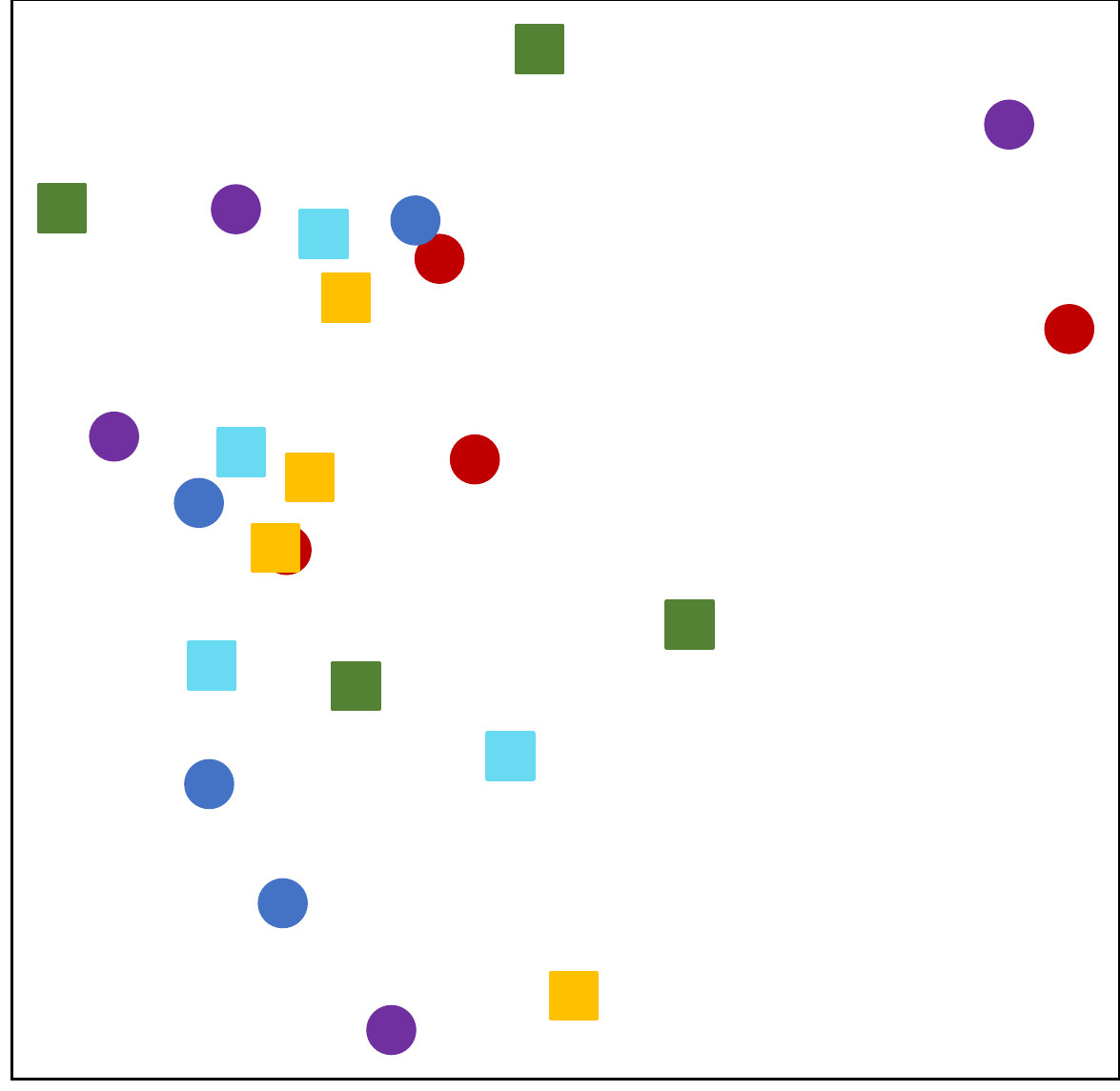}
}
\subfigure[HDGI]{
\includegraphics[width=0.18\linewidth]{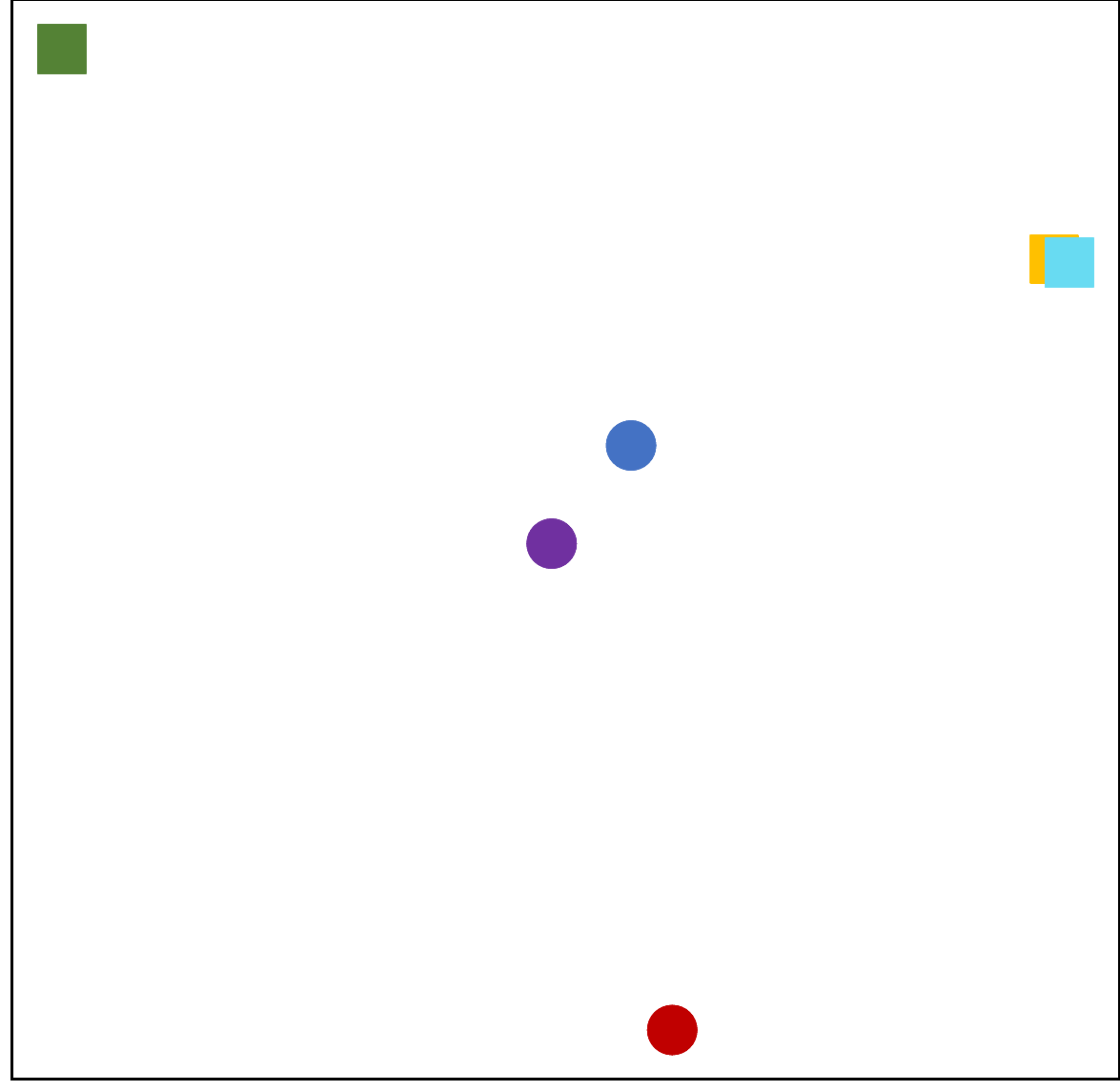}
}
\subfigure[HGT]{
\includegraphics[width=0.18\linewidth]{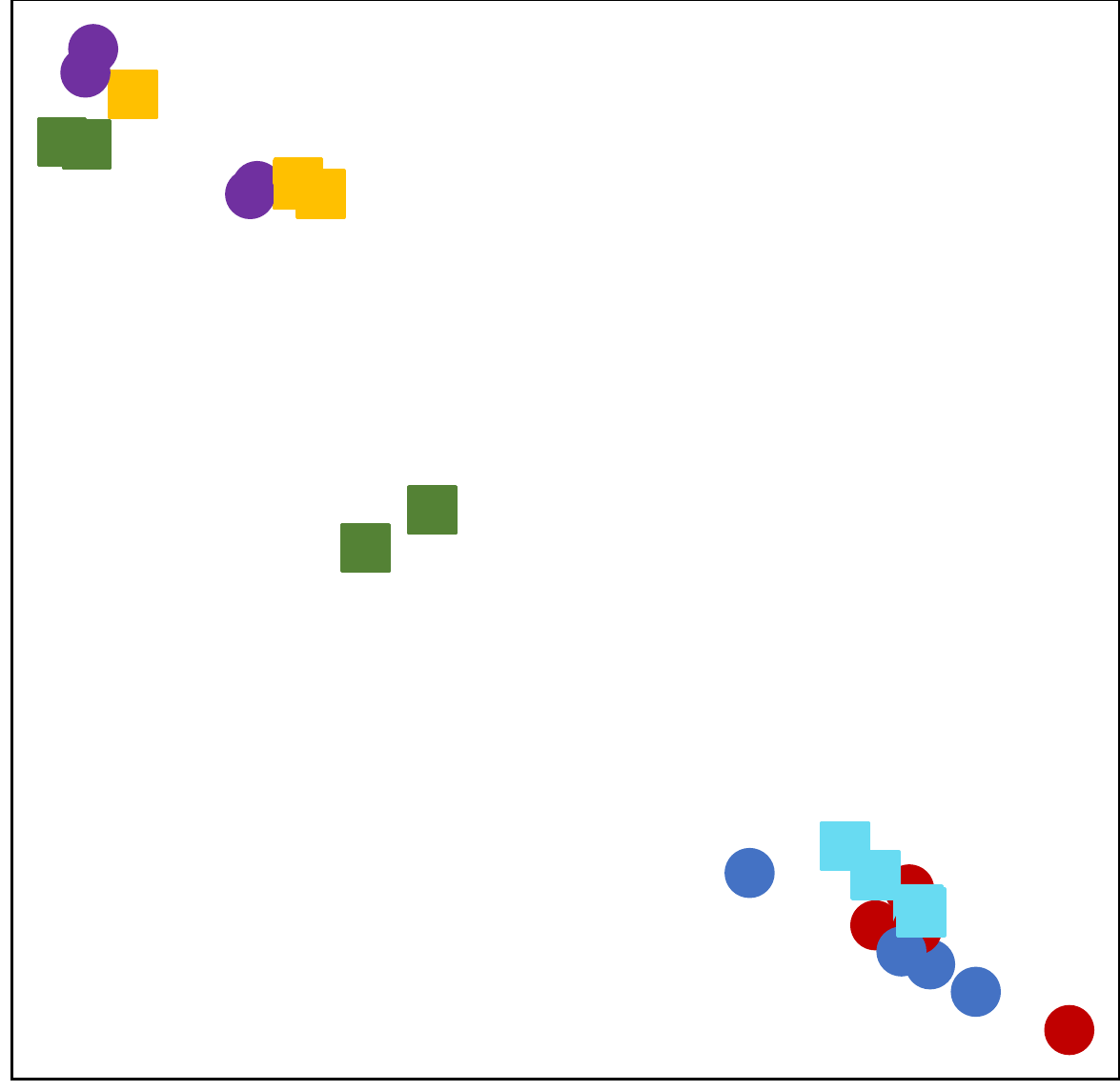}
}
\subfigure[HAWE (Ours)]{
\includegraphics[width=0.18\linewidth]{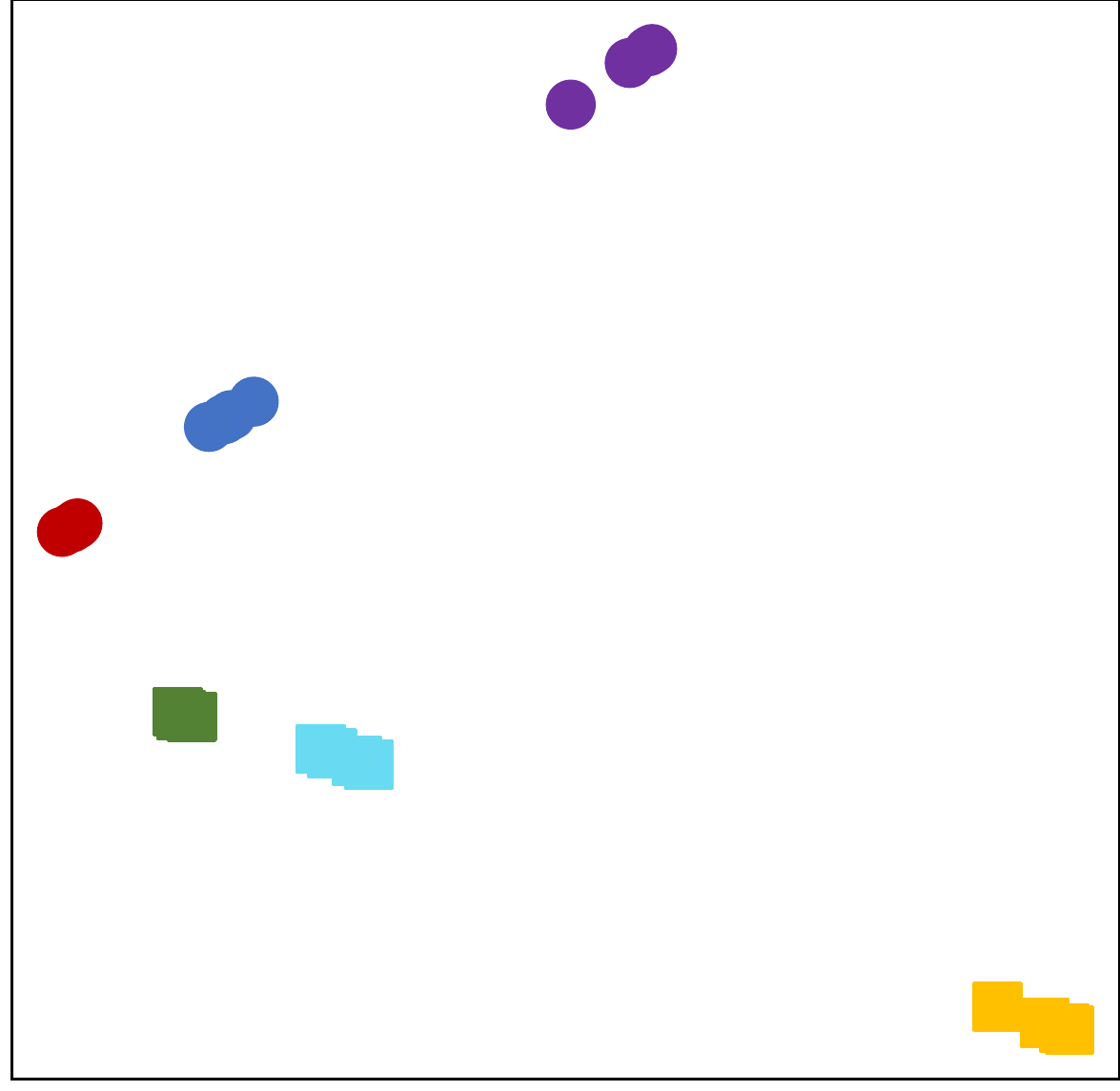}
}
\subfigure[CHAWE (Ours)]{
\includegraphics[width=0.18\linewidth]{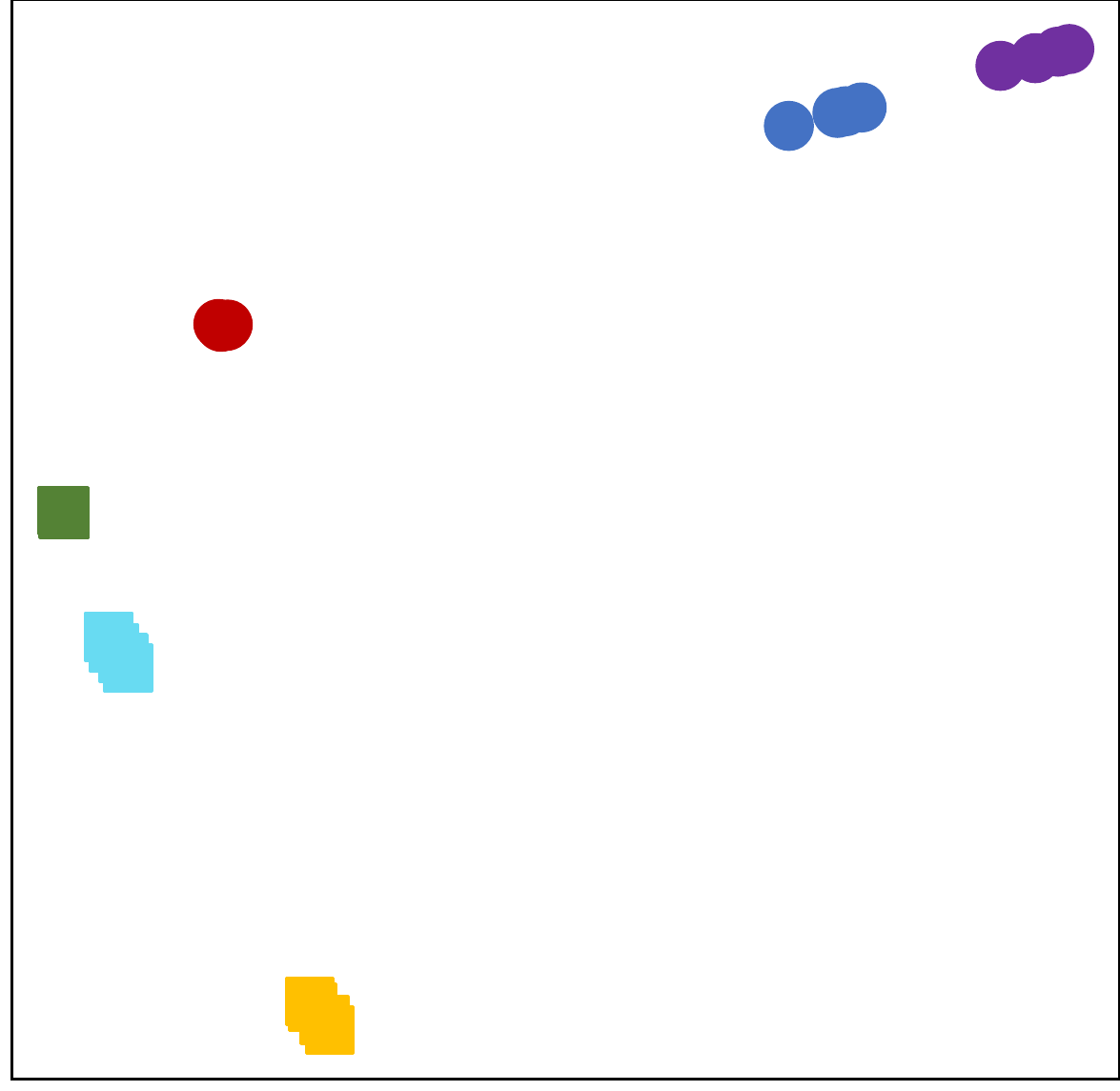}
}
\caption{2D-embedding visualization for all the baselines and our methods on the synthetic 
heterogeneous pinwheel network. The point shapes denote node types and the point colors denote heterostructural patterns.}
\label{fig.2d-vis}
\end{figure*}

\subsection{Real-world Datasets for Benchmark}
Before our work, processed datasets for heterostructure learning are scarce. We construct several heterogeneous networks based on real-world datasets for building an benchmark on heterostructure learning, where the node labels indicate their structural roles. 

One is an air-traffic network constructed based on the data collected by OpenFlights\footnote{ https://openflights.org/data.html, accessed Oct. 2021.} having two types of nodes: airports (A) and countries (C).The two types of edges denote air routes between airports and in which countries the airports are situated, respectively. The original dataset misses some detailed information of many airports including their countries. We fill in the missed information manually by searching the correponding IATA or ICAO codes online. We divide the airports into two classes based on their availability of international flights.

The others are Stack Exchange Q\&A networks on different topics\footnote{ https://archive.org/download/stackexchange, accessed Oct. 2021.}. For simplicity,  we refer to each of them as a prefix 'SE' with a suffix of its topic abbreviation in the following paper. They all have three types of nodes: users (U), questions (Q), and answers (A). The three types of edges represent the user giving a question/answer and the answer answering a question, respectively. For each network, we group users based on their reputation (a measurement of how much the community trusts the user) and answers based on their scores (the difference between its upvotes and downvotes) into balanced multiple classes respectively. There is also information about upvote and downvote counts for each user, which we use for more detailed demonstration in the experiment of similarity search. See Table \ref{tab.dataset} for some detailed statistics.

\subsection{Baseline Methods}
For comprehensively understanding the essence of our methods, we compare them with both homogeneous and heterogeneous network embedding methods. The homogeneous NE methods are as follows: 
\begin{itemize}

\item \textbf{DeepWalk} \cite{perozzi2014deepwalk} treats a network as a document and the nodes as words. It leverages random walks to extract the contexts of each node and applies a language model to generate embeddings. 

\item \textbf{LINE} \cite{tang2015line} learn embeddings by reconstructing the first-order and second-order proximities between nodes. 

\item \textbf{RolX} \cite{henderson2012rolx} learn non-negative role-based embeddings by factorizing the effective structural feature matrix generated by ReFeX \cite{henderson2011s}.

\item \textbf{Struc2vec} \cite{ribeiro2017struc2vec} constructs a multi-layer complete graph on the nodes of the original network based on computed pair-wise structural similarities. The similar mechanism used by DeepWalk is then applied. 

\item \textbf{GraphWave} \cite{donnat2018learning}
leverages heat wavelet diffusion patterns and learns structural embeddings via empirical characteristic functions of the wavelet coefficient distributions.

\item \textbf{Role2vec} \cite{ahmed2020role}
first assigns roles to nodes based on higher-order features and then applies random walk-based embedding method in which it replaces node Ids with roles.  

\item \textbf{GraphSTONE} \cite{long2020graph} leverages anonymous walks to capture structural patterns and gives a graph LDA model to capture the structural topics of each node. A two-view graph convolutional layer is designed to fusing both structural similarity and proximity into embeddings.
\end{itemize}

When we apply these homogeneous NE methods, we ignore the node types in the input heterogeneous networks. The heterogeneous NE methods are as follows: 
\begin{itemize}
\item \textbf{HIN2vec} \cite{fu2017hin2vec} uses random walks to generate node sequences. It learns the relations between nodes by predicting the meta-paths occurring in the sequences and cover the corresponding nodes.
\item \textbf{TransE} \cite{bordes2013translating} learns embeddings by translating each triplet (an typed edge and its two endpoints) into embedding distance calculation. 
\item \textbf{Node2bits} \cite{jin2019node2bits} designs biased random walks to aggregate neighbors' structures of each node and use a hashing method to generate embeddings. 
\item \textbf{R-GCN} \cite{schlichtkrull2018modeling} uses multiple graph convolutional layers to adaptively learn the corresponding type of relations among nodes.
\item \textbf{HDGI} \cite{ren2020heterogeneous} is an unsupervised method inheriting the basic architecture of HAN \cite{wang2019heterogeneous}, an attentive GNN model passing massages based on meta-paths. 
\item \textbf{HGT} \cite{hu2020heterogeneous} model the heterogeneity among nodes and edges with attention mechanisms and generate type-specific embeddings. For each pair of nodes, it tries to preserve the triplet of two node types and the edge type to train the embeddings.
\end{itemize}

\begin{table*}[!t]
\centering
\caption{Average accuracy of user classification on the Stack Exchange networks.}
\begin{tabular}{cccccccccccc}
\toprule
\textbf{Dataset} & SE-Anime & SE-Beer & SE-CG & SE-Chem & SE-CSE & SE-Engr & SE-FIT & SE-HWR & SE-IOT & SE-Latin & SE-Movie \\
\midrule
\textbf{DeepWalk} & 0.3611 & 0.4031 & 0.2895 & 0.2557 & 0.3967 & 0.2573 & 0.4105 & 0.3691 & 0.3663 & 0.3458 & 0.3859\\
\textbf{LINE} & 0.4772
 & 0.4574 & 0.3427 & 0.3532 & 0.4593 & 0.3201 & 0.4418 & 0.4209 & 0.4348 & 0.4191 & 0.4408
\\
\textbf{RolX} & 0.4755
 & 0.5563 & 0.3699 & 0.3825 & 0.4937 & 0.3354 & 0.5169 & 0.4426 & 0.3374 & 0.4581 & 0.4439
 \\
\textbf{struc2vec} & 0.4958  & 0.5055 & 0.3459 & 0.3574 & 0.4798 & 0.3348 & 0.4838 & 0.4132 & 0.4133 & 0.4373 & 0.4646 \\
\textbf{GraphWave} & 0.3383
 & 0.3273 & 0.3301 & 0.3151 & 0.3567  & 0.3913 & 0.3935 & 0.4109 & 0.2901
 & 0.3608 & 0.3742
 \\
\textbf{role2vec} & 0.4146
 & 0.4015 & 0.3536 & 0.3119 & 0.3644 & 0.4107 & 0.4399 & 0.4389 & 0.3463 & 0.4085
 & 0.4087
 \\
\textbf{GraphSTONE} & 0.4383
 & 0.4713  & 0.3581 & 0.3260 & 0.4572 & 0.3091 & 0.4694 & 0.4256 & 0.3124 & 0.4085 & 0.4099
 \\
\midrule
\textbf{HIN2vec} & 0.4865  & 0.5193 & 0.3521  & \textbf{0.3653} & 0.4770 & 0.3128 & 0.4976 & 0.4175 &  0.4428 & 0.4046 & 0.4519 \\
\textbf{TransE} & 0.4963
 & 0.5083 & 0.3483 & 0.3516 & 0.4453 & 0.3184 & 0.4811 & 0.4062 & 0.4093
 & 0.4388 & 0.4426
 \\
\textbf{node2bits} & 0.4924 & 0.5561 & 0.3544 & 0.3474 & 0.4851 & 0.3102 & 0.5112 & 0.4419 & 0.4374 & 0.4450 &  \textbf{0.4905}\\
\textbf{R-GCN} & 0.3608
 & 0.3346 & 0.2630 & 0.2646 & 0.3639 & 0.2574 & 0.3523 & 0.3426 & 0.3597
 & 0.3579 & 0.3412
\\
\textbf{HDGI} & 0.4935 & 0.5489 & 0.3573 & 0.3651 & 0.4775 & 0.3921 & 0.4927 & 0.4128 & 0.4243 & 0.4279 & 0.4609
\\
\textbf{HGT} & 0.4814
& 0.5242 & 0.3522 & 0.3473 & 0.4637 & 0.3094 & 0.4725 & 0.4259 & 0.4179
 & 0.4391 & 0.4517\\
\midrule
\textbf{HAWE} (ours) & \textbf{0.5307*} & \textbf{0.5954*} & \textbf{0.4166*} & 0.3615 & \textbf{0.4947} & \textbf{0.4152} & \textbf{0.5673} & \textbf{0.5033*} & \textbf{0.5138} & \textbf{0.5147} & \textbf{0.5028*}\\
\textbf{CHAWE} (ours) & \textbf{0.5177} & \textbf{0.5927} & \textbf{0.4069} & \textbf{0.4231*} & \textbf{0.5333*} & \textbf{0.4273*} & \textbf{0.5705*} & \textbf{0.4927} & \textbf{0.5270*} & \textbf{0.5292*} & 0.4868\\
\bottomrule
\end{tabular}

\label{tab.classification-user}
\end{table*}

\begin{table*}[!t]
\centering
\caption{Average accuracy of answer classification on the Stack Exchange networks.}
\begin{tabular}{cccccccccccc}
\toprule
\textbf{Dataset} & SE-Anime & SE-Beer & SE-CG & SE-Chem & SE-CSE & SE-Engr & SE-FIT & SE-HWR & SE-IOT & SE-Latin &SE-Movie\\
\midrule
\textbf{DeepWalk} & 0.3781 & 0.3048 & 0.4617 & 0.4102 & 0.3089 & 0.3385 & 0.2838 & 0.3712  & 0.5033 & 0.3674
 & 0.4025\\
\textbf{LINE} & 0.3348
 & 0.2581 & 0.3624 & 0.3358 &  0.2485 & 0.2573 & 0.2080 & 0.2581  & 0.3306 & 0.2531 & 0.3516
\\
\textbf{RolX} & 0.3753 & 0.3251 & 0.4175 & 0.4196 & 0.3083 & 0.3154 & 0.2889 & 0.3343 & 0.4457 & 0.3371 & 0.3924
\\
\textbf{struc2vec} & 0.3715 & 0.2905 & 0.3701 & 0.3732 & 0.2835 & 0.2911 & 0.2333 & 0.3078 & 0.3918  & 0.2819 & 0.3548 \\
\textbf{GraphWave} & 0.3056
 & 0.2222 & 0.3627 & 0.3043 & 0.2347  & 0.3252 & \textbf{0.3468} & 0.2721 & 0.3887 & 0.3095 & 0.3533
\\
\textbf{role2vec} & 0.3619
 & 0.3122 & 0.4727 & 0.3451 & 0.2832 & 0.3541 & \textbf{0.3678*} & 0.4039 & 0.5428 & 0.3664 & 0.3538
\\
\textbf{GraphSTONE} & 0.3846
 & 0.3043 & 0.4109 & 0.4054 & 0.2808 & 0.3073 & 0.2494 & 0.3792 & 0.4561 & 0.3439 & 0.4007
\\
\midrule
\textbf{HIN2vec} & 0.3775 & 0.3202 & 0.4528 & 0.4298 & 0.3223 & 0.3455 & 0.3075 & 0.3794 & 0.5140 & 0.3664 & 0.4093\\
\textbf{TransE} & 0.3672
 & 0.3098 & 0.4022 & 0.4177 & 0.3111
 & 0.2883 & 0.2613 & 0.3067 & 0.4483 & 0.3426 & 0.3802
\\
\textbf{node2bits} & 0.3963 & 0.3294 & 0.4213 & 0.4281 & 0.3172 & 0.3240 & 0.2914 & 0.3387  & 0.4414 & 0.3252  & 0.3745\\
\textbf{R-GCN} & 0.3445 & 0.2538 & 0.3458 & 0.3489 & 0.2649 & 0.2494 & 0.2227 & 0.2585  & 0.3974 & 0.3103 & 0.3346
\\
\textbf{HDGI} & 0.2978 & 0.3298 & 0.3254 & 0.3307 & 0.2461 & 0.2147 & 0.2398 & 0.2718 & 0.3623 & 0.2586 & 0.3757
\\
\textbf{HGT} & 0.3818
 & 0.3165 & 0.3923 & 0.4259 & 0.2890 & 0.2897 & 0.2373 & 0.3056 & 0.4194
 & 0.3181 & 0.3609\\
\midrule
\textbf{HAWE} (ours) & \textbf{0.4106} & \textbf{0.3788} & \textbf{0.5143} & \textbf{0.4336} & \textbf{0.3264} & \textbf{0.3823*} & 0.3370 & \textbf{0.4707*}  & \textbf{0.6249*} & \textbf{0.3863*} & \textbf{0.4159*}\\
\textbf{CHAWE} (ours) & \textbf{0.4194*} & \textbf{0.3805*}  & \textbf{0.5250*} & \textbf{0.4468*} & \textbf{0.3505*} & \textbf{0.3770} & 0.3342 & \textbf{0.4562}  & \textbf{0.6233} & \textbf{0.3844} & \textbf{0.4131}\\
\bottomrule
\end{tabular}

\label{tab.classification-answer}
\end{table*}

Note that RolX, struc2vec, GraphWave, role2vec, GraphSTONE and node2bits are designed for structure learning. On all experiments, the parameters of these baseline methods are finely tuned. For HDGI, we employ predefined meta-paths to build typed adjacency matrices. Specifically, we use \{ABA, ABBA, BAB, BAAB\} (A for circle and B for square) on heterogeneous pinwheel network, \{AA, ACA\} on Air-traffic network. On Stack Exchange networks, we apply \{UAU, UQU, UAQU\} for user classification and \{AUA, AQA, AQUA\} for answer classifcation.

\subsection{Model Configuration}
Except in parameter sensitivity analysis, we do the following configuration for both HAWE and CHAWE. We set sample size $T = 1024$ and window size $\Delta = 5$ on all the networks. For user classification on Stack Exchange networks, the walk length $L$ is set to $4$, while in other situations it is set to $6$. Embeddings are trained via stochastic gradient descent for $100$ epochs.

\subsection{2D-Visualization}
We employ our methods and most baseline methods on the synthetic heterogeneous pinwheel network shown in Fig. \ref{fig.pinwheel}(b) and generate 2-D node embeddings. For GraphWave whose embedding dimension cannot be changed and node2bits whose hashing process of  is invalid with too small embedding dimension, we generate higher-dimenional embeddings (100-D for GraphWave and 64-D for node2bits) and transform them into 2-D space via Principal Component Analysis (PCA). The 2-D visualization results are shown in Fig. \ref{fig.2d-vis}.

\begin{figure*}[!t]
\centering
\subfigure[The earliest user]{
\includegraphics[width=0.30\textwidth]{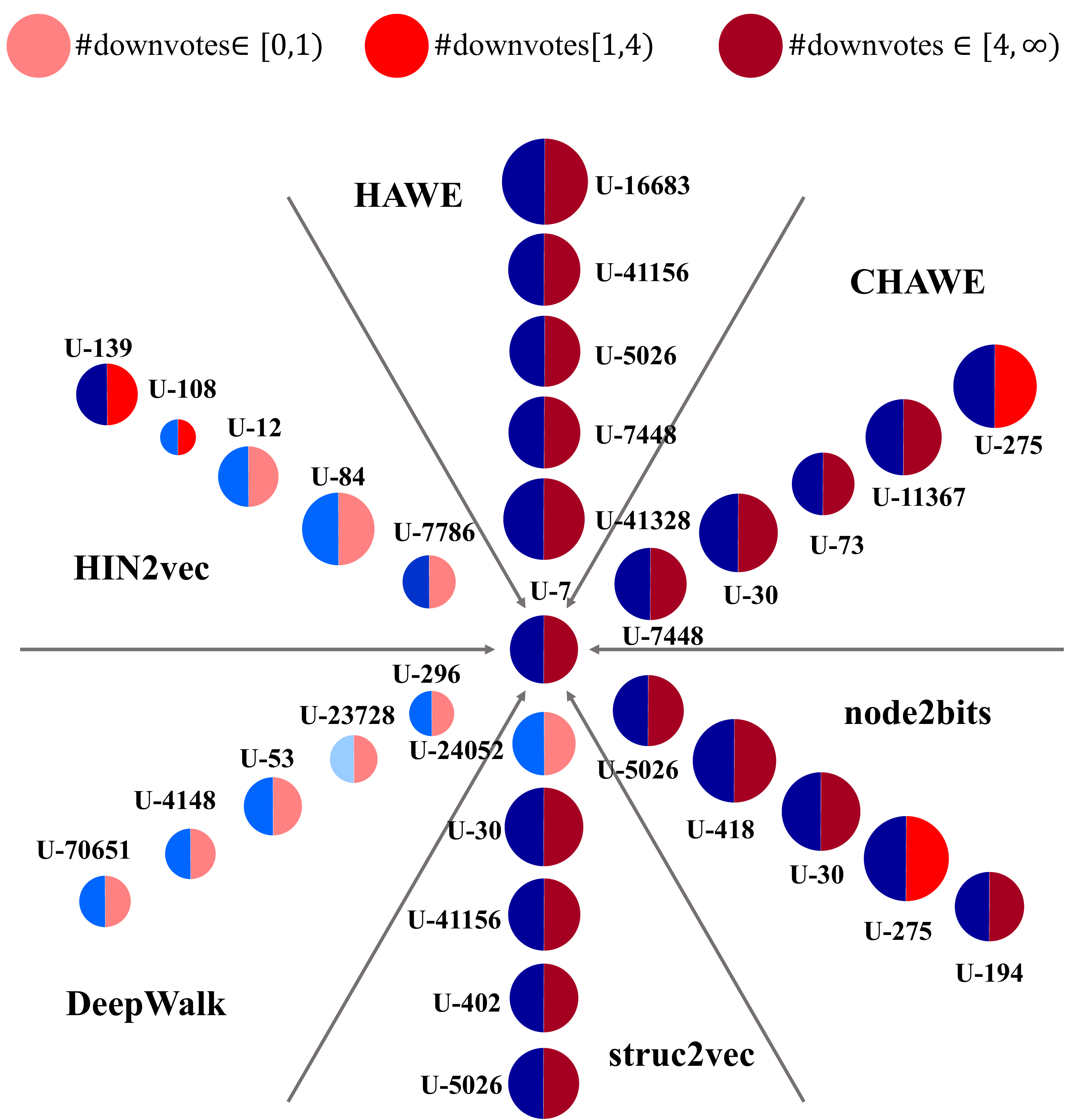}
}
\subfigure[The latest user]{
\includegraphics[width=0.30\textwidth]{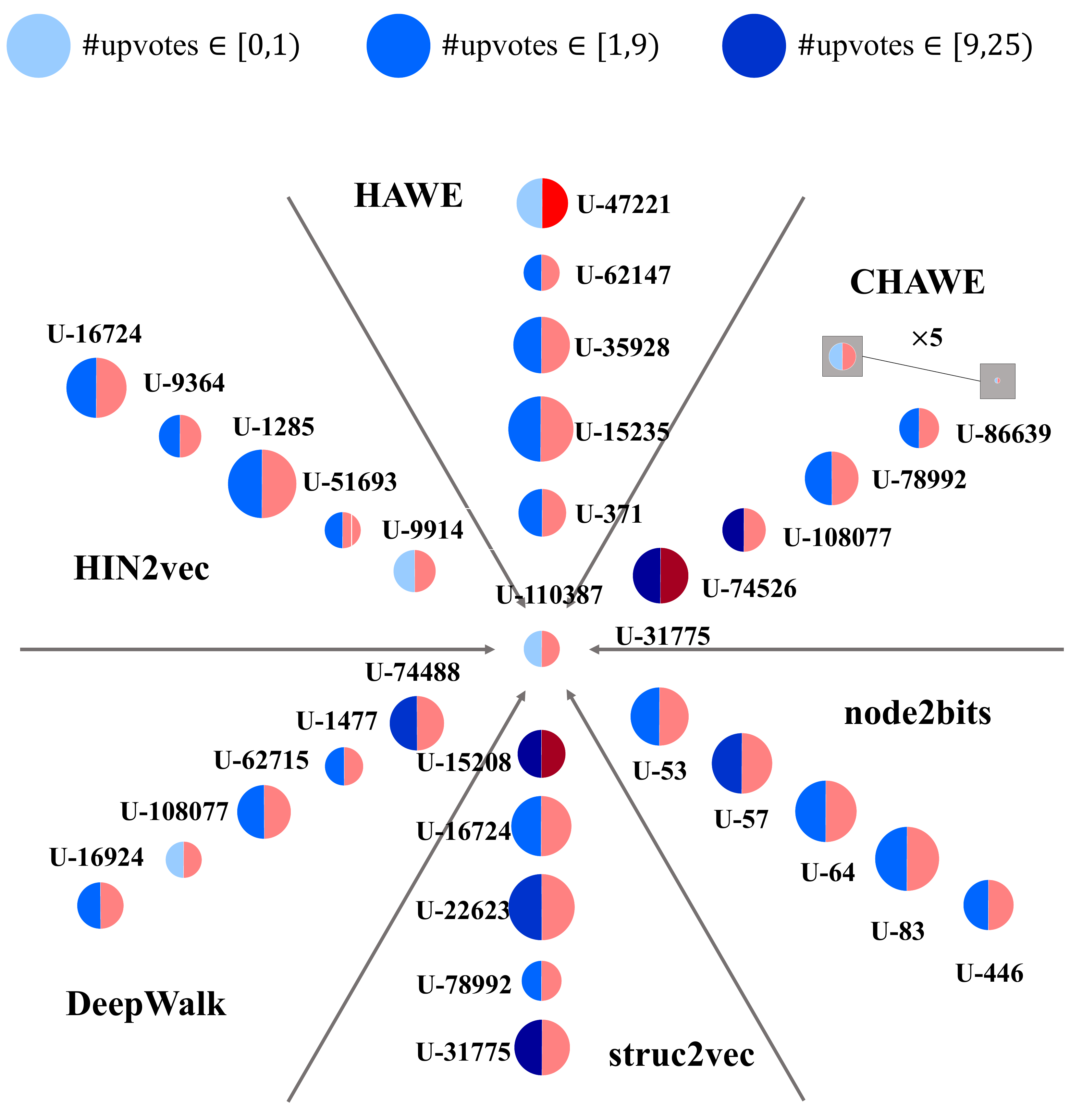}
}
\subfigure[The user with the highest reputation]{
\includegraphics[width=0.30\textwidth]{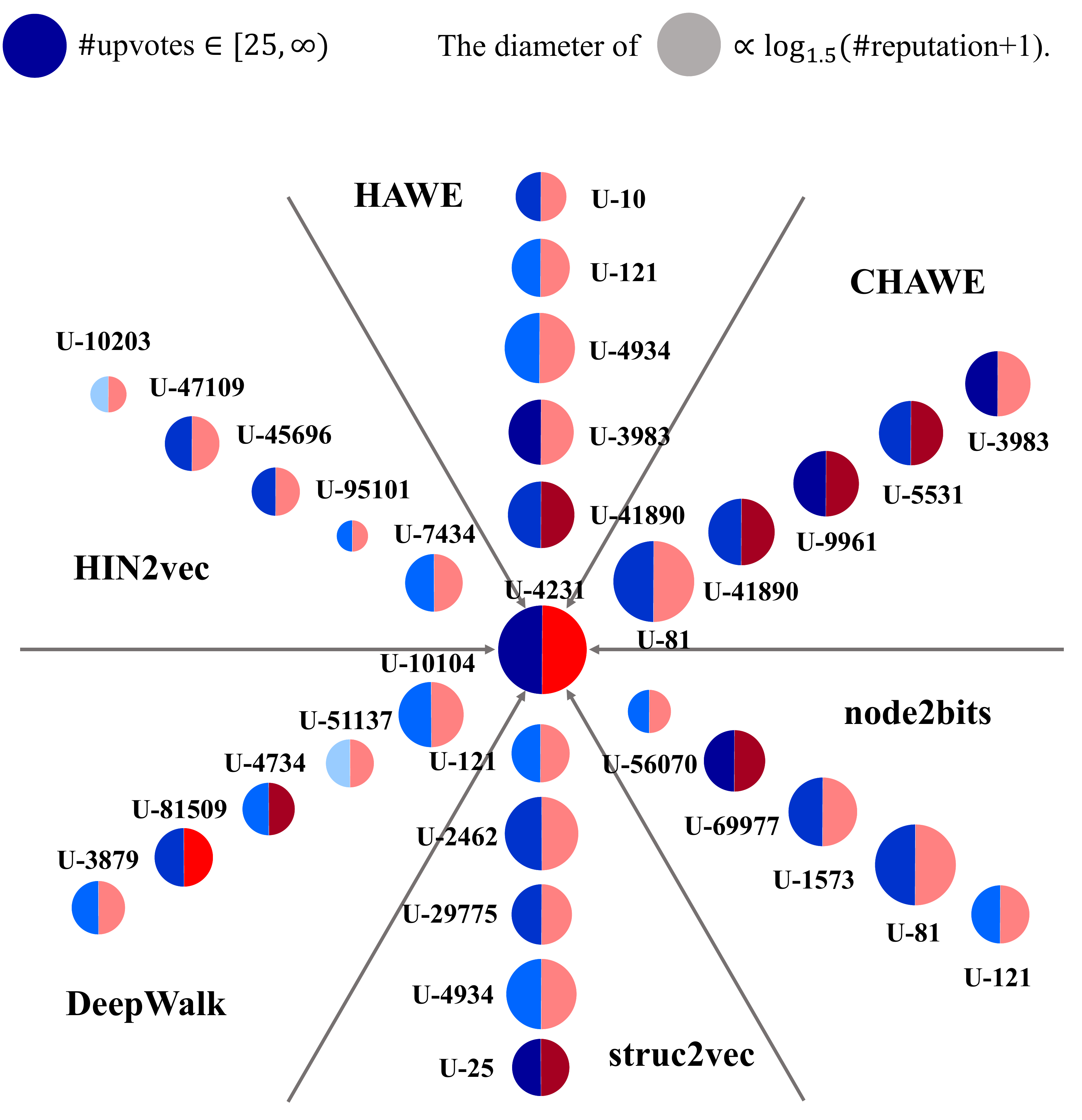}
}
\subfigure[The user with the lowest reputation]{
\includegraphics[width=0.30\textwidth]{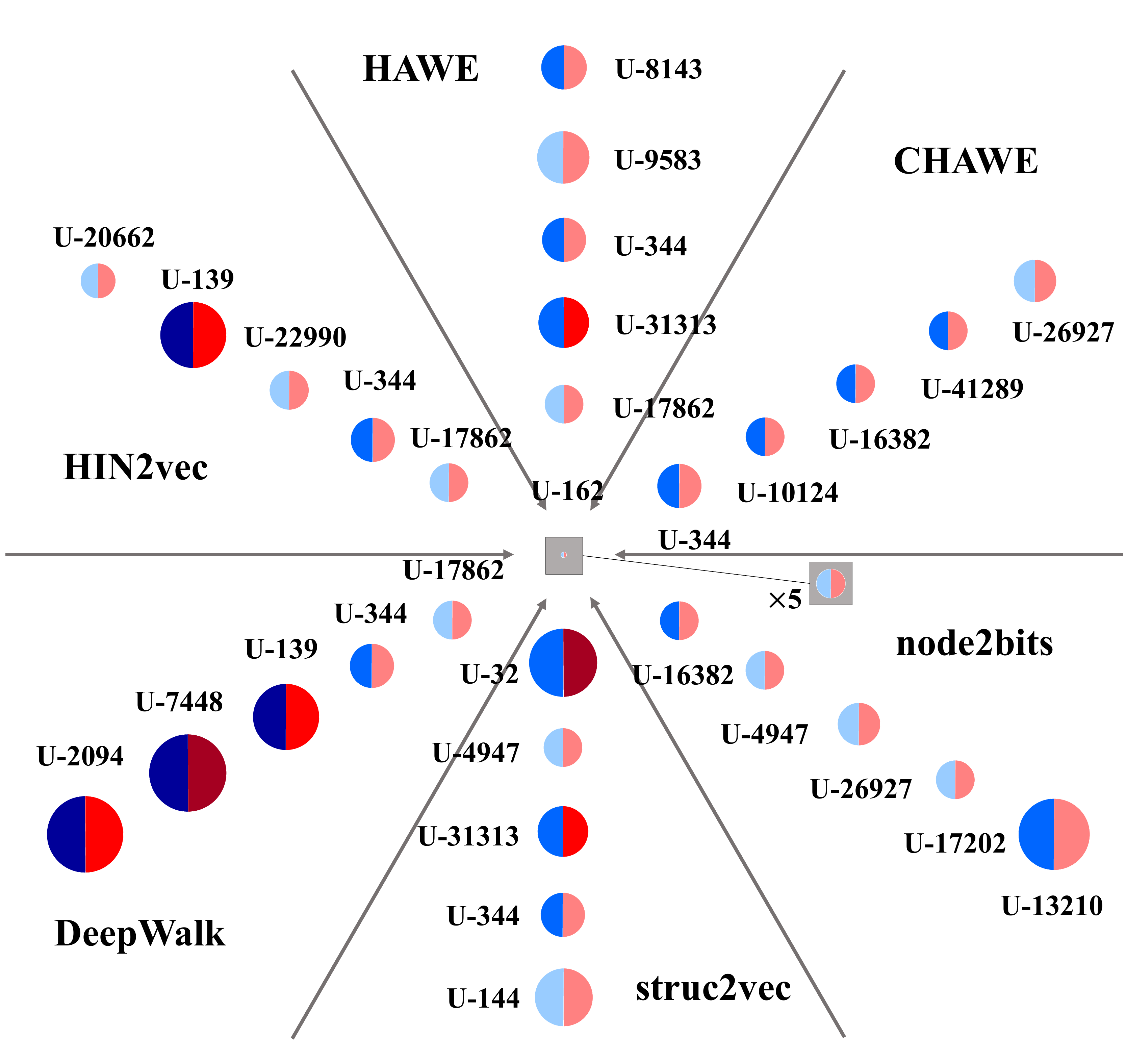}
}
\subfigure[The user with the most upvotes]{
\includegraphics[width=0.30\textwidth]{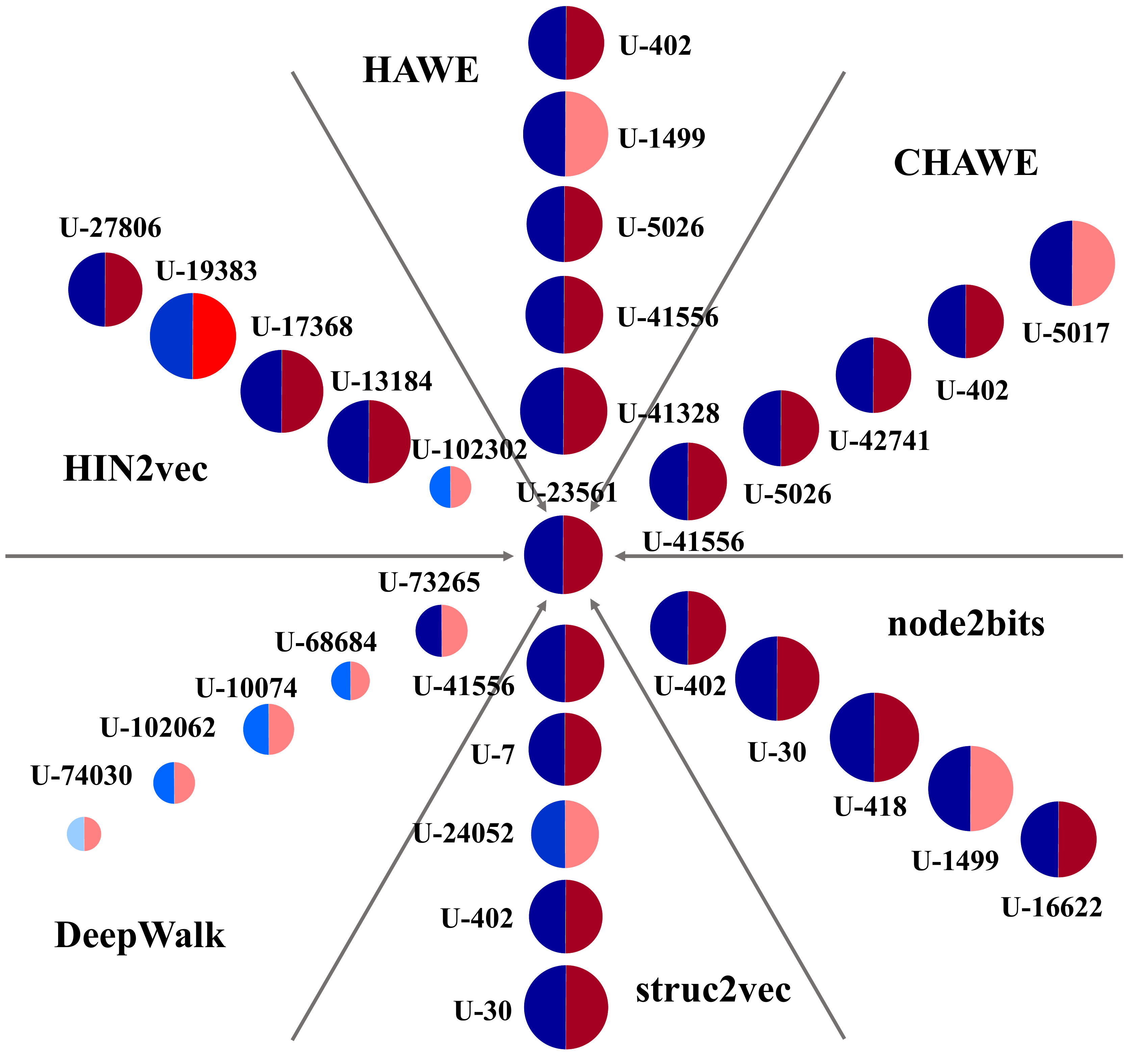}
}
\subfigure[The user with the most downvotes]{
\includegraphics[width=0.30\textwidth]{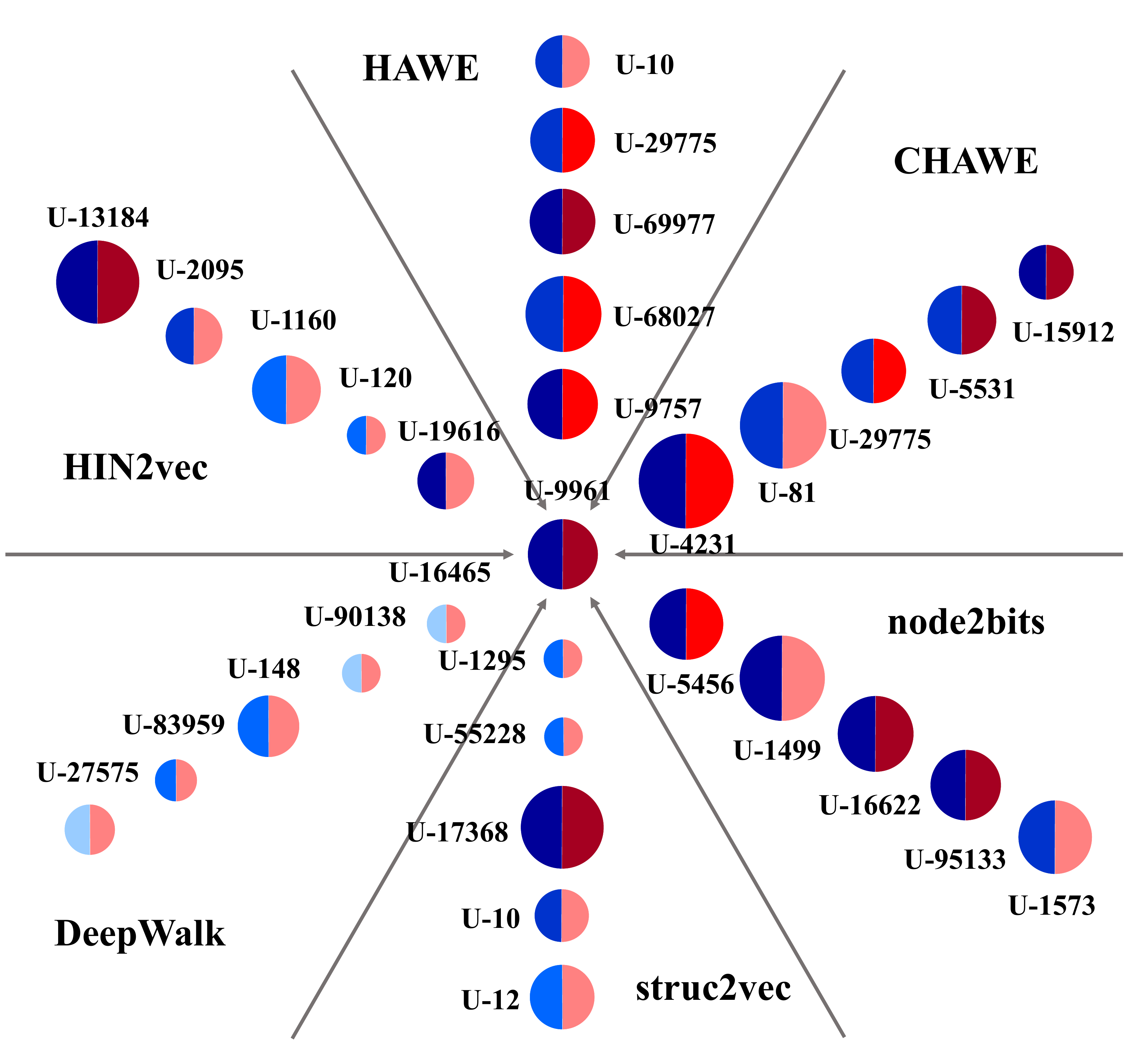}
}
\caption{The characteristic visualization of the closest 5 users in embedding space for some specific users. Circle size denotes the user reputation and color depth denote the upvote/downvote count of the user.}
\label{fig.case}
\end{figure*}

The nature of some methods is intuitively shown: DeepWalk, LINE, HIN2vec, TransE and HGT make embeddings of neighbors close while RolX, struc2vec and node2bits groups the nodes having the same neighborhood structures. 
We can observe that almost all the baseline methods cannot distinguish heterostrutures.  HDGI can gathers all nodes of the same role at one point because the role can be distinguished via the predefined meta-paths. For example, the blue nodes have no neighbors through the meta-path ABBA. But HDGI does not truly capture the heterostructures as it does not show the relation between nodes having the same underlying structures (e.g., the red and purple green). As we argued, node2bits is the only baseline method that captures both the type and structure of the nodes that make up the heterostructures. However, the its hashing process scatters the node in the same heterogeneous roles.
Both HAWE and CHAWE can effectively distinguish all the $6$ heterogeneous structural roles. And they do much better than node2bits as they make the nodes in the same role closer with lower-dimensional embeddings. And they more clearly show relations between the nodes having the same underlying patterns through their relative positions. The two node types are vertically distributed while the three underlying patterns are horizontally distributed in the same order. CHAWE groups the nodes in each roles closer than HAWE because of the fuzziness design of CHAW. Thus, HAW and CHAW do capture heterostructures and the embedding model can preserve them into representations.

\newcommand{\tabincell}[2]{\begin{tabular}
{@{}#1@{}}#2\end{tabular}}
\begin{table}[!t]
\centering
\caption{Average accuracy of airport classification \\ on the original and modified Air-traffic networks.}
\begin{tabular}{ccc}
\toprule
\textbf{Dataset} & \textbf{\tabincell{c}{Air-traffic \\(Original)}} & \textbf{\tabincell{c}{Air-traffic \\(Modified)}}\\
\midrule
\textbf{DeepWalk} & 0.8450 & 0.8269 \\
\textbf{LINE} & 0.8116 & 0.8002 \\
\textbf{RolX} & 0.8716
 & 0.8829
 \\
\textbf{struc2vec} & 0.8372 & 0.8333 \\
\textbf{GraphWave} & 0.8145 & 0.8258 \\
\textbf{role2vec} & 0.5348 & 0.5462 \\
\textbf{GraphSTONE} & 0.8288 & 0.8273 \\
\textbf{HIN2vec} & \textbf{0.9129} & 0.9051 \\
\textbf{TransE} & 0.8499
 & 0.8501 \\
\textbf{node2bits} & 0.8607 & 0.8601 \\
\textbf{R-GCN} & 0.5249 & 0.5227 \\
\textbf{HDGI} & 0.8365
 & 0.8523 \\
\textbf{HGT} & 0.8961 & 0.8841 \\
\textbf{HAWE} (ours) & \textbf{0.9242*} & \textbf{0.9351*} \\
\textbf{CHAWE} (ours) & 0.9001 & \textbf{0.9158} \\
\bottomrule
\end{tabular}

\label{tab.classification}
\end{table}
\subsection{Heterogeneous Structural Role Classification}
We conduct heterogeneous structural role classification experiments on real-world networks. Specifically, we apply all methods on these networks and generate 128-D embeddings (100-D for GraphWave).  For each method, we take $70\%$ of generated embeddings as the training set to train a Logistic Regression classifier. Then we apply the trained classifiers on test sets, i.e.,  the other $30\%$ embeddings, and calculate the classification accuracy. We repeat above process for 50 times and report average accuracy in Table \ref{tab.classification-user}-\ref{tab.classification}. On each task, the top 2 results are bold while the best one is marked with the symbol \textbf{*}.

\begin{figure*}[!t]
\centering
\includegraphics[width=0.97\linewidth]{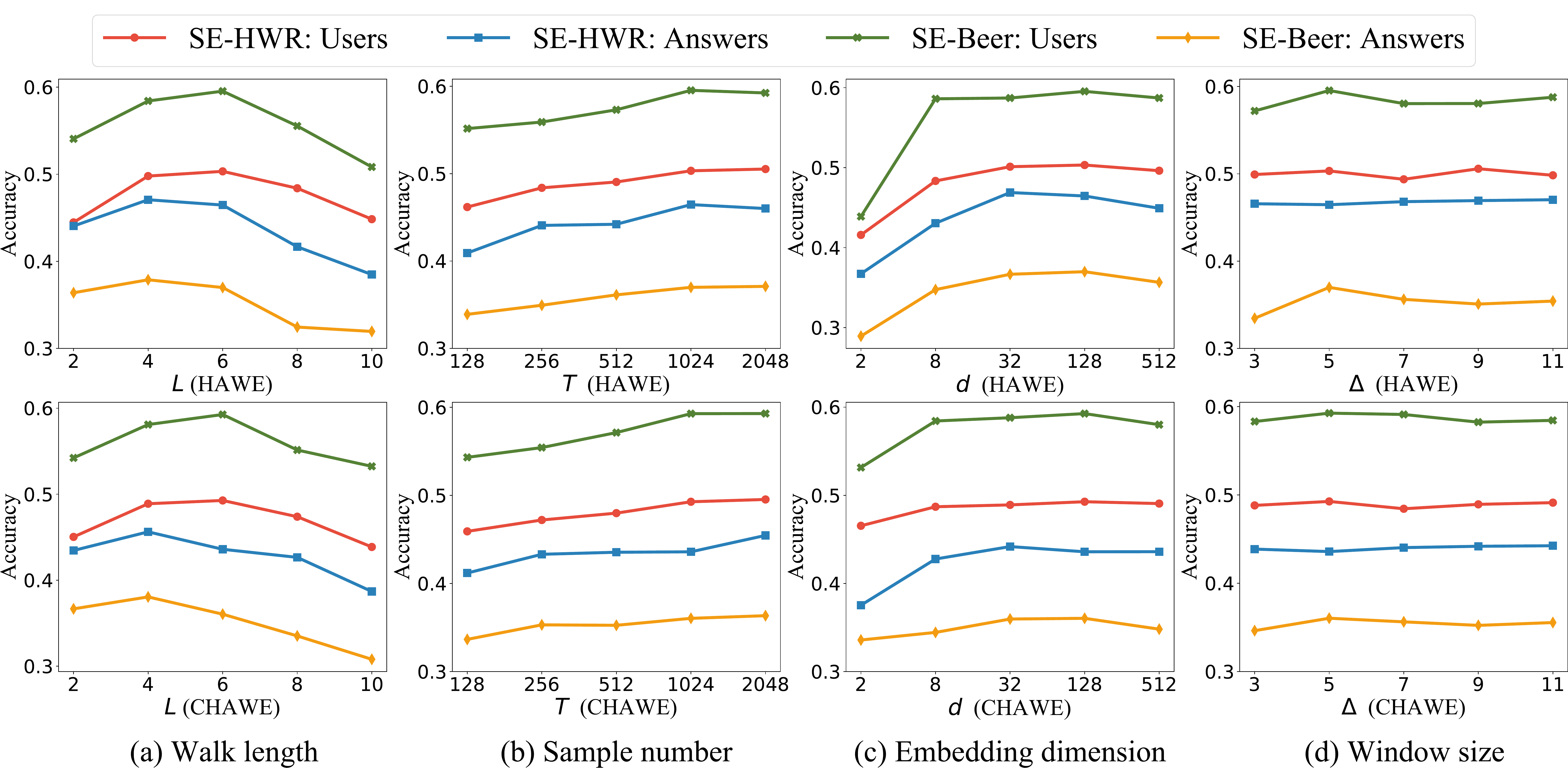}
\caption{Sensitivity analysis results on parameters including: walk length $L$, sample number $T$, embedding dimension $d$ and window size $\Delta$.}
\label{fig.para}
\end{figure*}

In the air-traffic (Original) network, every domestic airport and all of its neighbor airports are connected to the same country node, while the neighbor airports of an international airports may belong to different countries. Thus, the essential task is to detect a 4-path subgraph in which two connected airports are connected to different countries respectively. As designed for capturing this kind of heterostructures, our HAWE unsurprisingly performs the best. And it is ineluctable for CHAWE to get lower results than HAWE because of the need for delicate detection. We can observe from reported results that most methods achieve great performance. This is because the domestic airports gather as communities so that methods capturing proximities can also detection them. And international airports usually have higher degree than domestic airports, which makes methods learning structures work. For verification, we modify the original air-traffic network by deleting $50\%$ of edges among the domestic airports and redo the experiments on it. As expected, baseline methods capturing proximities get worse performance in the modified airport network, and performance of those capturing structures almost stays unchanged. Accuracy of HAWE and CHAWE increases because the heterostructural traits of the two kinds of airports becomes more prominent.

On each Stack Exchange network, we classify users and answers respectively. On user classification, RolX, struc2vec and node2bits outperform other baseline methods over most networks. These methods leverage statistical features which are strongly correlated with user reputation such as node degrees. The other structural embedding methods including GraphWave, role2vec and GraphSTONE do not achieve competitive results due to the 
failure of their structure capture mechanisms (e.g., wavelet coefficient distributions and AWs) on heterogeneous networks. In contrast, the feature-based methods show no superiority compared with the other baseline methods on answer classification. This is because the score of the answer is influenced by both the reputation of the respondent and the quality of the answer itself. The latter is not related to the neighborhood heterostructures of answers and cannot be discriminated by all the compared methods. What's more, high-quality answers are usually provided by the users having high reputation. Thus, there is proximity between high-score answers, which leads to the competitive results of some proximity-based methods such as DeepWalk and HIN2vec. However, in almost all networks, our methods get the top $2$ results on both user classification and answer classification because of their superiority in capturing heterostructures.


\begin{table}[!t]
\centering
\caption{Characteristics of some specific users in SE-Chem.}
\begin{tabular}{cccc}
\toprule
\textbf{User ID} & \textbf{$\#$ reputation} & \textbf{$\#$ upvotes} & \textbf{$\#$ downvotes} \\
\midrule
U-7    &   5,868   &   26   &   9   \\
U-110387 &   101   &   0   &   0   \\
U-4231    &   79,505   &   132   &   2   \\
U-162    &   1   &   0   &   0  \\
U-23561    &   11,596   &   1,594   &   123   \\
U-9961    &   4,202   &   297   &   297   \\
\bottomrule
\end{tabular}

\label{tab.users}
\end{table}

\subsection{Similarity Search}
We further design a top-k similarity search experiment to illustrate more details. We compute the Euclidean distances between user embeddings generated on SE-Chem network by each method. Then we retrieve $5$ users whose corresponding embeddings are the closest to that of a target user. 

In Fig. \ref{fig.case}, we show the search results of HAWE, CHAWE, node2bits, struc2vec, HIN2vec and DeepWalk on $6$ specific users. Each user is displayed as a circle consisting of two semicircles. The circle size denotes user reputation and the semicircle color depth denotes the upvote/downvote count range of a user. The ranges are divided in a balanced manner. We also show the user IDs recorded in the raw data. The target users include: (a) the user (U-7) creating the account the earliest in 2012; (b) the user (U-110387) creating the account the latest in 2021; (c) the user (U-4231) having the highest reputation; (d) the user (U-162) having the lowest reputation; (e) the user (U-23561) having the most upvotes; (f) the user (U-9961) having the most downvotes. The characteristics of the them are provided in Table \ref{tab.users}. 

In the ideal case, the users retrieved by a method good at learning heterostructures should have similar circle sizes and semicircle color depths to the target user. With this criterion, we can observe that HAWE and CHAWE are in the top tier of performance. And CHAWE is better than HAWE due to the fuzzy design of CHAWs on similar heterostructures. Methods capturing structures, i.e., node2bit and struc2vec achieve the level of the second tier. HIN2vec and DeepWalk often retrieve users with relatively low reputation which verifies their nature of capturing proximities. Therefore, the results of similarity search comprehensively show the superiority of our HAWE and CHAWE in learning heterostructures 
at the micro level.

\subsection{Parameter Sensitivity Analysis}
In this part, we study how the important parameters including walk length $L$, sample number $T$, embedding dimension $d$ and window size $\Delta$ influence the effectiveness of HAWE and CHAWE. Specifically, we employ both user and answer classification on SE-Beer and SE-HWR networks with one parameter changing and the other parameters fixed. When the parameters are fixed, we set $L = 6$, $T = 1024$, $d = 128$ and $\Delta = 5$ . The results are demonstrated in Fig. \ref{fig.para}. 

We can observe that each parameter affects the two methods in the same way: 
\begin{enumerate}
\item When $L$ is the variable, the accuracy increases first and then decreases with the growth of $L$. This is because the sampled (C)HAWs have trouble in capture heterostructures when they are too short. When walk length is too long, the captured heterostructures are overly diverse and much more samples are needed. The extreme point of $L$ on answer classification is smaller than that on user classification. Because the roles (i.e., quality) of answers is more depend more on the roles of close nodes than the roles of users.
\item The accuracy increases with the growth of $L$. The more walks are sampled, the more delicate distributions of (C)HAWs are estimated and the model are trained with more training samples (Eq. (\ref{eq:prediction})).
\item As embedding dimension $d$ increases, the accuracy increases first for more heterostructure information preserved. Then it almost remains constant or decreases slightly as the excessive embedding dimension is redundant for preserving heterostructure information and leads to overfitting problem.
\item The window size $\Delta$ has little affect on the performance when it is large enough. When $\Delta \ge 5$, the information of (C)HAWs sampled by the sliding window is enough for predicting an unknown (C)HAW in the same context. 
\end{enumerate}

\begin{figure}[!t]
\centering
\includegraphics[width=0.95\linewidth]{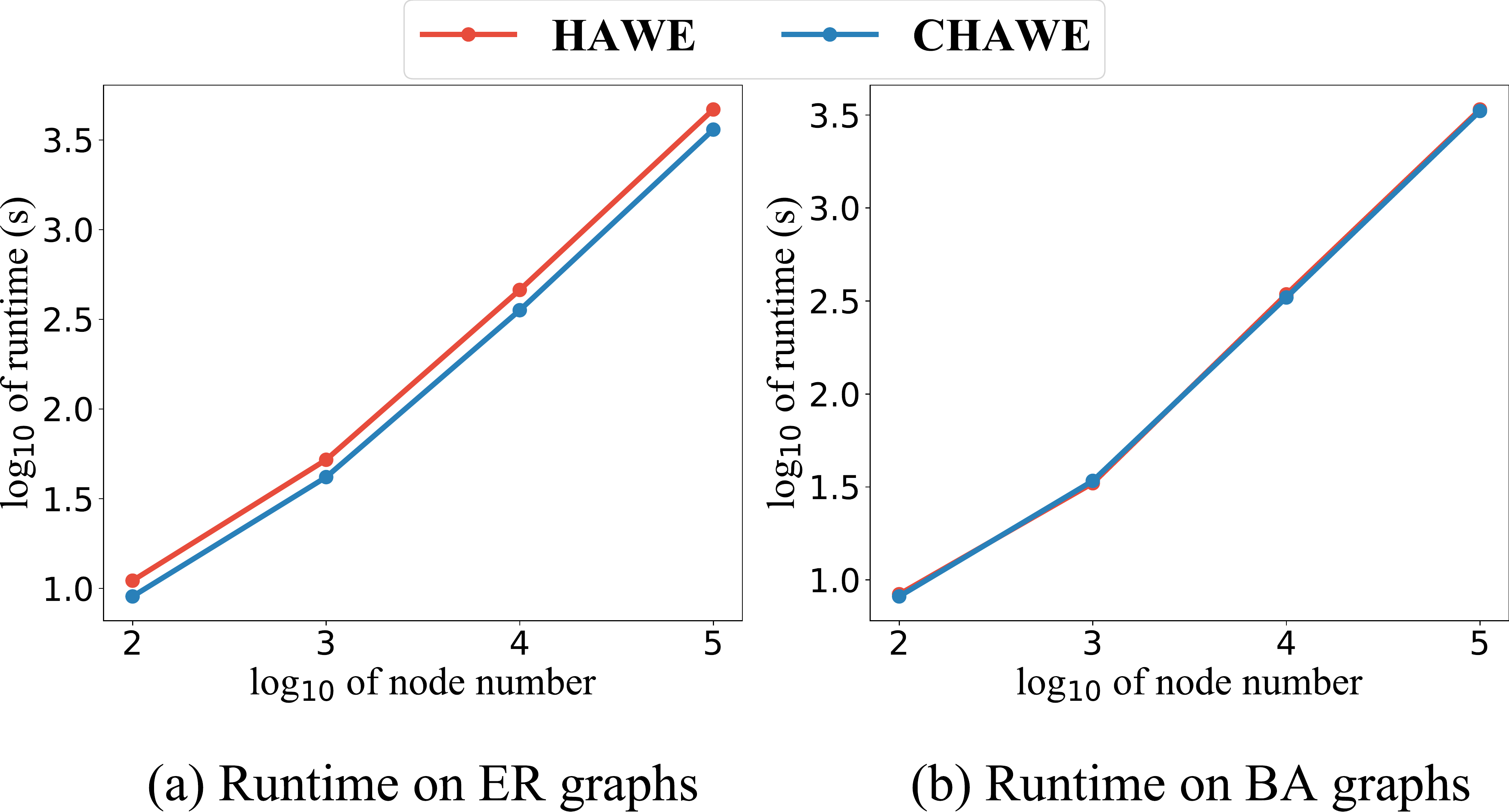}
\caption{Runtime of HAWE and CHAWE on ER and BA graphs.}
\label{fig.time}
\end{figure}

\subsection{Runtime Analysis}
To evaluate the scalability of our proposed methods, we generate two series of synthetic graphs of which the node numbers range from $100$ to $100000$ via Erdos-Renyi (ER) model \cite{gilbert1959random} and Barabási-Albert (BA) model \cite{barabasi1999emergence}, respectively. For ER model, we set the probability of linking two arbitrary nodes to $10/N$ so that the average node degree is approximately fixed to $5$. For BA model, we  set the number of edge linked from a new node to existing nodes to $1$ so that the generated graphs are trees. The nodes in these networks are aligned to $2$ types randomly. We train HAWE and CHAWE on these synthetic graphs for $100$ epochs with all the parameters fixed: $L = 6$, $T = 1024$, $d = 128$ and $\Delta = 5$. We run the methods on each network for $5$ times and illustrate the average runtime results in Fig. \ref{fig.time}. Intuitively, the cost time of both HAWE and CHAWE is linear with the edge number of the network, which verifies our conclusion derived in Section \ref{sec:time}. On ER graphs, HAWE costs more time than CHAWE. On BA graphs, the runtime of the two methods are almost the same. This is because the HAWE generates larger lexicon $\mathcal{L}$ than CHAWE on ER graphs, while the generated lexicons of both methods on BA graphs have small and similar sizes due to the tree structure. In real-world networks, since there are lots of loops, CHAWE is also more efficient than HAWE.

\section{Related Work}

\subsection{Homogeneous Structural Embedding}
Almost all the existing algorithms of structural node embedding are designed for homogeneous networks \cite{Jin2021Toward,jiao2021survey}. Usually, these work use a strategy of extracting structural properties such as extracting features \cite{henderson2012rolx}, estimating wavelet distribution \cite{donnat2018learning} and applying graph kernel method \cite{nikolentzos2019learning} before mapping them to vector space \cite{jiao2021survey}. To transform captured structural traits into embeddings, the early studies such as RolX \cite{henderson2012rolx} and GLRD \cite{gilpin2013guided} directly factorize processed structural feature matrix. Later a few works extend random walk-based methods (e.g., DeepWalk \cite{perozzi2014deepwalk}). Struc2vec \cite{ribeiro2017struc2vec} achieves structural similar nodes tending to occur in the same walk by constructing a new random walk graph based on node degrees. RiWalk \cite{xuewei2019riwalk} make structural similar nodes have similar walks starting from themselves by proposing a new node labeling method. Recently, using deep learning to generate structural embeddings has 
attracted more attention. DRNE \cite{tu2018deep} captures regular equivalence by aggregating degrees of node neighbors with Long Short Term Memory \cite{hochreiter1997long}. CNESE \cite{guo2021learning} applies variational auto-encoder \cite{Kingma2014} to learn stochastic equivalence by reconstructing the distributions of discrete graph curvature. GraLSP\cite{jin2020gralsp} and  GraphSTONE\cite{long2020graph} use anonymous walks to capture neighborhood structures and learn embeddings through graph neural network architecture.

\subsection{Heterogeneous Network Embedding}
NE methods on heterogeneous network are mainly devised for capturing the node proximity \cite{yang2020heterogeneous}. The earliest works of heterogeneous NE study knowledge graphs. These methods, such as TransE \cite{bordes2013translating}
, train embeddings by learning a scoring function measuring how accurate embeddings representing the heterogeneous triplets. Later a number of methods extending previous proximity-preserving homogeneous NE methods. Metapath2vec \cite{dong2017metapath2vec} and HIN2vec \cite{fu2017hin2vec} extend DeepWalk \cite{perozzi2014deepwalk} and preserve node proximity with random walks based on meta-paths. PTE \cite{tang2015pte} and AspEm \cite{shi2018aspem} extend LINE \cite{tang2015line} by designing edge-type based and aspect-based proximities, respectively. Recently, deep-learning based methods have become the main branch of heterogeneous NE. For example, R-GCN \cite{schlichtkrull2018modeling} applies multiple GCNs \cite{kipf2017semisupervised} to learn the edge heterogeneity. HDGI \cite{ren2020heterogeneous}, an unsupervised version of HAN\cite{wang2019heterogeneous} on which the contrastive architecture of DGI \cite{velickovic2019deep} is grafted. It uses both meta-path based neighbor-level and semantic-level attention mechanisms to aggregate deep heterogeneous proximity information. HGT \cite{hu2020heterogeneous} uses a series of attention mechanisms to reconstruct the typed node pair in each triplet . HetSAGNN \cite{hong2020attention} designs a type-aware attention layer to aggregate the embeddings of different types of nodes locating in different space. It trains the embeddings through multi-task learning.

To our best knowledge, exsiting heterogeneous 
NE methods except node2bits\cite{jin2019node2bits} cannot capture the highly diverse heterostructures. Node2bits extracts structural features for each type of neighbors sampled by random walks and generate embeddings via hashing methods. Its design for efficiency leads to low fineness on capturing heterostructures. 

\section{Conclusion and Future Works}
Learning the complex heterostructures, i.e., the combinations of the node types and underlying structures, is a critical but underappreciated problem in the field of heterogeneous network embedding.
In this paper, we make a first attempt at NE on heterostructure learning. We propose the promising HAW which has theoretically guaranteed ability to distinguish heterostructures and its more practical variant CHAW. We take advantages of HAWs and CHAWs by sampling them as the context of each node's heterostructure theme and provide an embedding method HAWE and its variant CHAWE by imitating a language model. Finally, we provide the first benchmark on learning heterogeneous structural roles. A number of datasets and tasks are proposed. As expected, our methods show amazing performance on them.

Although the proposed methods have a great ability to capture heterostructures, they are not perfect. Theoretically, the methods should achieve better performance when sampling longer (C)HAWs. But sampling overlong samples is not feasible due to the extremely sparsity of the samples. What's more, many sampled (C)HAWs are different but corresponded to the same induced heterogeneous subgraph. In other words, there are many synonyms in the generated heterostructure contexts. 
It is the inevitable result of the sampling strategy using (C)HAWs. But if we can reduce the number of synonyms, the effectiveness of our methods will be improved. We leave these problems as our future works.

\section*{Acknowledgment}
This work is supported by the National Natural Science Foundation of China (61902278).

\ifCLASSOPTIONcaptionsoff
  \newpage
\fi



\bibliographystyle{IEEEtran}
\bibliography{tkde}

\begin{thebibliography}{10}
\providecommand{\url}[1]{#1}
\csname url@samestyle\endcsname
\providecommand{\newblock}{\relax}
\providecommand{\bibinfo}[2]{#2}
\providecommand{\BIBentrySTDinterwordspacing}{\spaceskip=0pt\relax}
\providecommand{\BIBentryALTinterwordstretchfactor}{4}
\providecommand{\BIBentryALTinterwordspacing}{\spaceskip=\fontdimen2\font plus
\BIBentryALTinterwordstretchfactor\fontdimen3\font minus
  \fontdimen4\font\relax}
\providecommand{\BIBforeignlanguage}[2]{{%
\expandafter\ifx\csname l@#1\endcsname\relax
\typeout{** WARNING: IEEEtran.bst: No hyphenation pattern has been}%
\typeout{** loaded for the language `#1'. Using the pattern for}%
\typeout{** the default language instead.}%
\else
\language=\csname l@#1\endcsname
\fi
#2}}
\providecommand{\BIBdecl}{\relax}
\BIBdecl

\bibitem{zhang2018network}
D.~Zhang, J.~Yin, X.~Zhu, and C.~Zhang, ``Network representation learning: A
  survey,'' \emph{IEEE Transactions on Big Data}, vol.~6, no.~1, pp. 3--28,
  2018.

\bibitem{boguna2021network}
M.~Boguna, I.~Bonamassa, M.~De~Domenico, S.~Havlin, D.~Krioukov, and M.~{\'A}.
  Serrano, ``Network geometry,'' \emph{Nature Reviews Physics}, vol.~3, no.~2,
  pp. 114--135, 2021.

\bibitem{ribeiro2017struc2vec}
L.~F. Ribeiro, P.~H. Saverese, and D.~R. Figueiredo, ``struc2vec: Learning node
  representations from structural identity,'' in \emph{Proceedings of the 23rd
  ACM SIGKDD International Conference on Knowledge Discovery and Data Mining},
  2017, pp. 385--394.

\bibitem{jiao2021temporal}
P.~Jiao, X.~Guo, X.~Jing, D.~He, H.~Wu, S.~Pan, M.~Gong, and W.~Wang,
  ``Temporal network embedding for link prediction via vae joint attention
  mechanism,'' \emph{IEEE Transactions on Neural Networks and Learning
  Systems}, pp. 1--14, 2021.

\bibitem{zhang2019iteratively}
W.~Zhang, B.~Paudel, L.~Wang, J.~Chen, H.~Zhu, W.~Zhang, A.~Bernstein, and
  H.~Chen, ``Iteratively learning embeddings and rules for knowledge graph
  reasoning,'' in \emph{The World Wide Web Conference}, 2019, pp. 2366--2377.

\bibitem{perozzi2014deepwalk}
B.~Perozzi, R.~Al-Rfou, and S.~Skiena, ``Deepwalk: Online learning of social
  representations,'' in \emph{Proceedings of the 20th ACM SIGKDD International
  Conference on Knowledge Discovery and Data Mining}, 2014, pp. 701--710.

\bibitem{rossi2020proximity}
R.~A. Rossi, D.~Jin, S.~Kim, N.~K. Ahmed, D.~Koutra, and J.~B. Lee, ``On
  proximity and structural role-based embeddings in networks: Misconceptions,
  techniques, and applications,'' \emph{ACM Transactions on Knowledge Discovery
  from Data}, vol.~14, no.~5, pp. 1--37, 2020.

\bibitem{jiao2021survey}
P.~Jiao, X.~Guo, T.~Pan, W.~Zhang, Y.~Pei, and L.~Pan, ``A survey on
  role-oriented network embedding,'' \emph{IEEE Transactions on Big Data}, pp.
  1--20, 2021.

\bibitem{rossi2014role}
R.~A. Rossi and N.~K. Ahmed, ``Role discovery in networks,'' \emph{IEEE
  Transactions on Knowledge and Data Engineering}, vol.~27, no.~4, pp.
  1112--1131, 2014.

\bibitem{guo2020role}
X.~Guo, W.~Zhang, W.~Wang, Y.~Yu, Y.~Wang, and P.~Jiao, ``Role-oriented graph
  auto-encoder guided by structural information,'' in \emph{International
  Conference on Database Systems for Advanced Applications}.\hskip 1em plus
  0.5em minus 0.4em\relax Springer, 2020, pp. 466--481.

\bibitem{rossi2020structural}
R.~A. Rossi, N.~K. Ahmed, E.~Koh, S.~Kim, A.~Rao, and Y.~Abbasi-Yadkori, ``A
  structural graph representation learning framework,'' in \emph{Proceedings of
  the 13th International Conference on Web Search and Data Mining}, 2020, pp.
  483--491.

\bibitem{ma2019riwalk}
X.~Ma, G.~Qin, Z.~Qiu, M.~Zheng, and Z.~Wang, ``Riwalk: Fast structural node
  embedding via role identification,'' in \emph{2019 IEEE International
  Conference on Data Mining}.\hskip 1em plus 0.5em minus 0.4em\relax IEEE,
  2019, pp. 478--487.

\bibitem{nikolentzos2019learning}
G.~Nikolentzos and M.~Vazirgiannis, ``Learning structural node representations
  using graph kernels,'' \emph{IEEE Transactions on Knowledge and Data
  Engineering}, 2019.

\bibitem{jin2020gralsp}
Y.~Jin, G.~Song, and C.~Shi, ``Gralsp: Graph neural networks with local
  structural patterns,'' in \emph{Proceedings of the AAAI Conference on
  Artificial Intelligence}, vol.~34, 2020, pp. 4361--4368.

\bibitem{long2020graph}
Q.~Long, Y.~Jin, G.~Song, Y.~Li, and W.~Lin, ``Graph structural-topic neural
  network,'' in \emph{Proceedings of the 26th ACM SIGKDD International
  Conference on Knowledge Discovery \& Data Mining}, 2020, pp. 1065--1073.

\bibitem{long2021theoretically}
Q.~Long, Y.~Jin, Y.~Wu, and G.~Song, ``Theoretically improving graph neural
  networks via anonymous walk graph kernels,'' in \emph{Proceedings of the Web
  Conference 2021}, 2021, pp. 1204--1214.

\bibitem{micali2016reconstructing}
S.~Micali and Z.~A. Zhu, ``Reconstructing markov processes from independent and
  anonymous experiments,'' \emph{Discrete Applied Mathematics}, vol. 200, pp.
  108--122, 2016.

\bibitem{yang2020heterogeneous}
C.~Yang, Y.~Xiao, Y.~Zhang, Y.~Sun, and J.~Han, ``Heterogeneous network
  representation learning: A unified framework with survey and benchmark,''
  \emph{IEEE Transactions on Knowledge and Data Engineering}, 2020.

\bibitem{dong2017metapath2vec}
Y.~Dong, N.~V. Chawla, and A.~Swami, ``metapath2vec: Scalable representation
  learning for heterogeneous networks,'' in \emph{Proceedings of the 23rd ACM
  SIGKDD International Conference on knowledge discovery and data mining},
  2017, pp. 135--144.

\bibitem{fu2017hin2vec}
T.-y. Fu, W.-C. Lee, and Z.~Lei, ``Hin2vec: Explore meta-paths in heterogeneous
  information networks for representation learning,'' in \emph{Proceedings of
  the 2017 ACM on Conference on Information and Knowledge Management}, 2017,
  pp. 1797--1806.

\bibitem{schlichtkrull2018modeling}
M.~Schlichtkrull, T.~N. Kipf, P.~Bloem, R.~Van Den~Berg, I.~Titov, and
  M.~Welling, ``Modeling relational data with graph convolutional networks,''
  in \emph{European Semantic Web Conference}.\hskip 1em plus 0.5em minus
  0.4em\relax Springer, 2018, pp. 593--607.

\bibitem{hong2020attention}
H.~Hong, H.~Guo, Y.~Lin, X.~Yang, Z.~Li, and J.~Ye, ``An attention-based graph
  neural network for heterogeneous structural learning,'' in \emph{Proceedings
  of the AAAI Conference on Artificial Intelligence}, vol.~34, 2020, pp.
  4132--4139.

\bibitem{hu2020heterogeneous}
Z.~Hu, Y.~Dong, K.~Wang, and Y.~Sun, ``Heterogeneous graph transformer,'' in
  \emph{Proceedings of The Web Conference 2020}, 2020, pp. 2704--2710.

\bibitem{jin2019node2bits}
D.~Jin, M.~Heimann, R.~Rossi, and D.~Koutra, ``node2bits: Compact time-and
  attribute-aware node representations,'' in \emph{ECML/PKDD European
  Conference on Principles and Practice of Knowledge Discovery in Databases},
  2019.

\bibitem{milo2002network}
R.~Milo, S.~Shen-Orr, S.~Itzkovitz, N.~Kashtan, D.~Chklovskii, and U.~Alon,
  ``Network motifs: simple building blocks of complex networks,''
  \emph{Science}, vol. 298, no. 5594, pp. 824--827, 2002.

\bibitem{gardner1978bells}
M.~Gardner, ``Bells-versatile numbers that can count partitions of a set,
  primes and even rhymes,'' \emph{Scientific American}, vol. 238, no.~5, pp.
  24--30, 1978.

\bibitem{rossi2020heterogeneous}
R.~A. Rossi, N.~K. Ahmed, A.~Carranza, D.~Arbour, A.~Rao, S.~Kim, and E.~Koh,
  ``Heterogeneous graphlets,'' \emph{ACM Transactions on Knowledge Discovery
  from Data (TKDD)}, vol.~15, no.~1, pp. 1--43, 2020.

\bibitem{ivanov2018anonymous}
S.~Ivanov and E.~Burnaev, ``Anonymous walk embeddings,'' in \emph{International
  Conference on Machine Learning}.\hskip 1em plus 0.5em minus 0.4em\relax PMLR,
  2018, pp. 2186--2195.

\bibitem{le2014distributed}
Q.~Le and T.~Mikolov, ``Distributed representations of sentences and
  documents,'' in \emph{International Conference on Machine Learning}.\hskip
  1em plus 0.5em minus 0.4em\relax PMLR, 2014, pp. 1188--1196.

\bibitem{XuHLJ19}
K.~Xu, W.~Hu, J.~Leskovec, and S.~Jegelka, ``How powerful are graph neural
  networks?'' in \emph{International Conference on Learning Representations},
  2019.

\bibitem{mikolov2013distributed}
T.~Mikolov, I.~Sutskever, K.~Chen, G.~S. Corrado, and J.~Dean, ``Distributed
  representations of words and phrases and their compositionality,'' in
  \emph{Advances in Neural Information Processing Systems}, 2013, pp.
  3111--3119.

\bibitem{tang2015line}
J.~Tang, M.~Qu, M.~Wang, M.~Zhang, J.~Yan, and Q.~Mei, ``Line: Large-scale
  information network embedding,'' in \emph{Proceedings of the 24th
  International Conference on World Wide Web}, 2015, pp. 1067--1077.

\bibitem{henderson2012rolx}
K.~Henderson, B.~Gallagher, T.~Eliassi-Rad, H.~Tong, S.~Basu, L.~Akoglu,
  D.~Koutra, C.~Faloutsos, and L.~Li, ``Rolx: structural role extraction \&
  mining in large graphs,'' in \emph{Proceedings of the 18th ACM SIGKDD
  International Conference on Knowledge Discovery and Data Mining}, 2012, pp.
  1231--1239.

\bibitem{henderson2011s}
K.~Henderson, B.~Gallagher, L.~Li, L.~Akoglu, T.~Eliassi-Rad, H.~Tong, and
  C.~Faloutsos, ``It's who you know: graph mining using recursive structural
  features,'' in \emph{Proceedings of the 17th ACM SIGKDD International
  Conference on Knowledge Discovery and Data Mining}, 2011, pp. 663--671.

\bibitem{donnat2018learning}
C.~Donnat, M.~Zitnik, D.~Hallac, and J.~Leskovec, ``Learning structural node
  embeddings via diffusion wavelets,'' in \emph{Proceedings of the 24th ACM
  SIGKDD International Conference on Knowledge Discovery \& Data Mining}, 2018,
  pp. 1320--1329.

\bibitem{ahmed2020role}
N.~Ahmed, R.~A. Rossi, J.~Lee, T.~Willke, R.~Zhou, X.~Kong, and H.~Eldardiry,
  ``Role-based graph embeddings,'' \emph{IEEE Transactions on Knowledge and
  Data Engineering}, 2020.

\bibitem{bordes2013translating}
A.~Bordes, N.~Usunier, A.~Garcia-Duran, J.~Weston, and O.~Yakhnenko,
  ``Translating embeddings for modeling multi-relational data,'' \emph{Advances
  in Neural Information Processing Systems}, vol.~26, 2013.

\bibitem{ren2020heterogeneous}
Y.~Ren and B.~Liu, ``Heterogeneous deep graph infomax,'' in \emph{Workshop of
  Deep Learning on Graphs: Methodologies and Applications co-located with the
  Thirty-Fourth AAAI Conference on Artificial Intelligence}, 2020.

\bibitem{wang2019heterogeneous}
X.~Wang, H.~Ji, C.~Shi, B.~Wang, Y.~Ye, P.~Cui, and P.~S. Yu, ``Heterogeneous
  graph attention network,'' in \emph{The World Wide Web Conference}, 2019, pp.
  2022--2032.

\bibitem{gilbert1959random}
E.~N. Gilbert, ``Random graphs,'' \emph{The Annals of Mathematical Statistics},
  vol.~30, no.~4, pp. 1141--1144, 1959.

\bibitem{barabasi1999emergence}
A.-L. Barab{\'a}si and R.~Albert, ``Emergence of scaling in random networks,''
  \emph{science}, vol. 286, no. 5439, pp. 509--512, 1999.

\bibitem{Jin2021Toward}
J.~Jin, M.~Heimann, D.~Jin, and D.~Koutra, ``Toward understanding and
  evaluating structural node embeddings,'' \emph{ACM Transactions on Knowledge
  Discovery from Data}, vol.~16, no.~3, 2021.

\bibitem{gilpin2013guided}
S.~Gilpin, T.~Eliassi-Rad, and I.~Davidson, ``Guided learning for role
  discovery (glrd) framework, algorithms, and applications,'' in
  \emph{Proceedings of the 19th ACM SIGKDD International Conference on
  Knowledge Discovery and Data Mining}, 2013, pp. 113--121.

\bibitem{xuewei2019riwalk}
M.~Xuewei, G.~Qin, Z.~Qiu, M.~Zheng, and Z.~Wang, ``Riwalk: Fast structural
  node embedding via role identification,'' in \emph{2019 IEEE International
  Conference on Data Mining (ICDM)}, 2019, pp. 478--487.

\bibitem{tu2018deep}
K.~Tu, P.~Cui, X.~Wang, P.~S. Yu, and W.~Zhu, ``Deep recursive network
  embedding with regular equivalence,'' in \emph{Proceedings of the 24th ACM
  SIGKDD International Conference on Knowledge Discovery and Data Mining},
  2018, pp. 2357--2366.

\bibitem{hochreiter1997long}
S.~Hochreiter and J.~Schmidhuber, ``Long short-term memory,'' \emph{Neural
  computation}, vol.~9, no.~8, pp. 1735--1780, 1997.

\bibitem{guo2021learning}
X.~Guo, Q.~Tian, W.~Zhang, W.~Wang, and P.~Jiao, ``Learning stochastic
  equivalence based on discrete ricci curvature,'' in \emph{30th International
  Joint Conference on Artificial Intelligence}, 2021.

\bibitem{Kingma2014}
D.~P. Kingma and M.~Welling, ``{Auto-encoding variational bayes},'' in
  \emph{International Conference on Learning Representations}, 2014.

\bibitem{tang2015pte}
J.~Tang, M.~Qu, and Q.~Mei, ``Pte: Predictive text embedding through
  large-scale heterogeneous text networks,'' in \emph{Proceedings of the 21th
  ACM SIGKDD International Conference on Knowledge Discovery and Data Mining},
  2015, pp. 1165--1174.

\bibitem{shi2018aspem}
Y.~Shi, H.~Gui, Q.~Zhu, L.~Kaplan, and J.~Han, ``Aspem: Embedding learning by
  aspects in heterogeneous information networks,'' in \emph{Proceedings of the
  2018 SIAM International Conference on Data Mining}.\hskip 1em plus 0.5em
  minus 0.4em\relax SIAM, 2018, pp. 144--152.

\bibitem{kipf2017semisupervised}
T.~N. Kipf and M.~Welling, ``Semi-supervised classification with graph
  convolutional networks,'' in \emph{International Conference on Learning
  Representations}, 2018.

\bibitem{velickovic2019deep}
P.~Velickovic, W.~Fedus, W.~L. Hamilton, P.~Li{\`o}, Y.~Bengio, and R.~D.
  Hjelm, ``Deep graph infomax.'' in \emph{International Conference on Learning
  Representations}, 2019.

\end{thebibliography}
%

%

\begin{IEEEbiography}[{\includegraphics[width=1in,height=1.25in,clip,keepaspectratio]{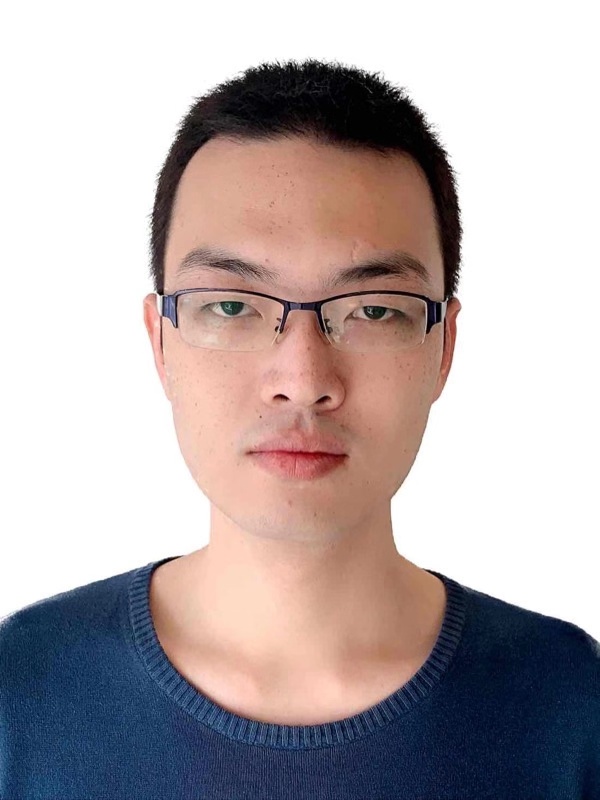}}]{Xuan Guo} is pursuing a doctoral degree at the College of Intelligence and Computing, Tianjin University, Tianjin, China. His current research interests include complex network analysis, role discovery,  network representation learning and network percolation model.
\end{IEEEbiography}

\begin{IEEEbiography}[{\includegraphics[width=1in, height=1.25in,clip,keepaspectratio]{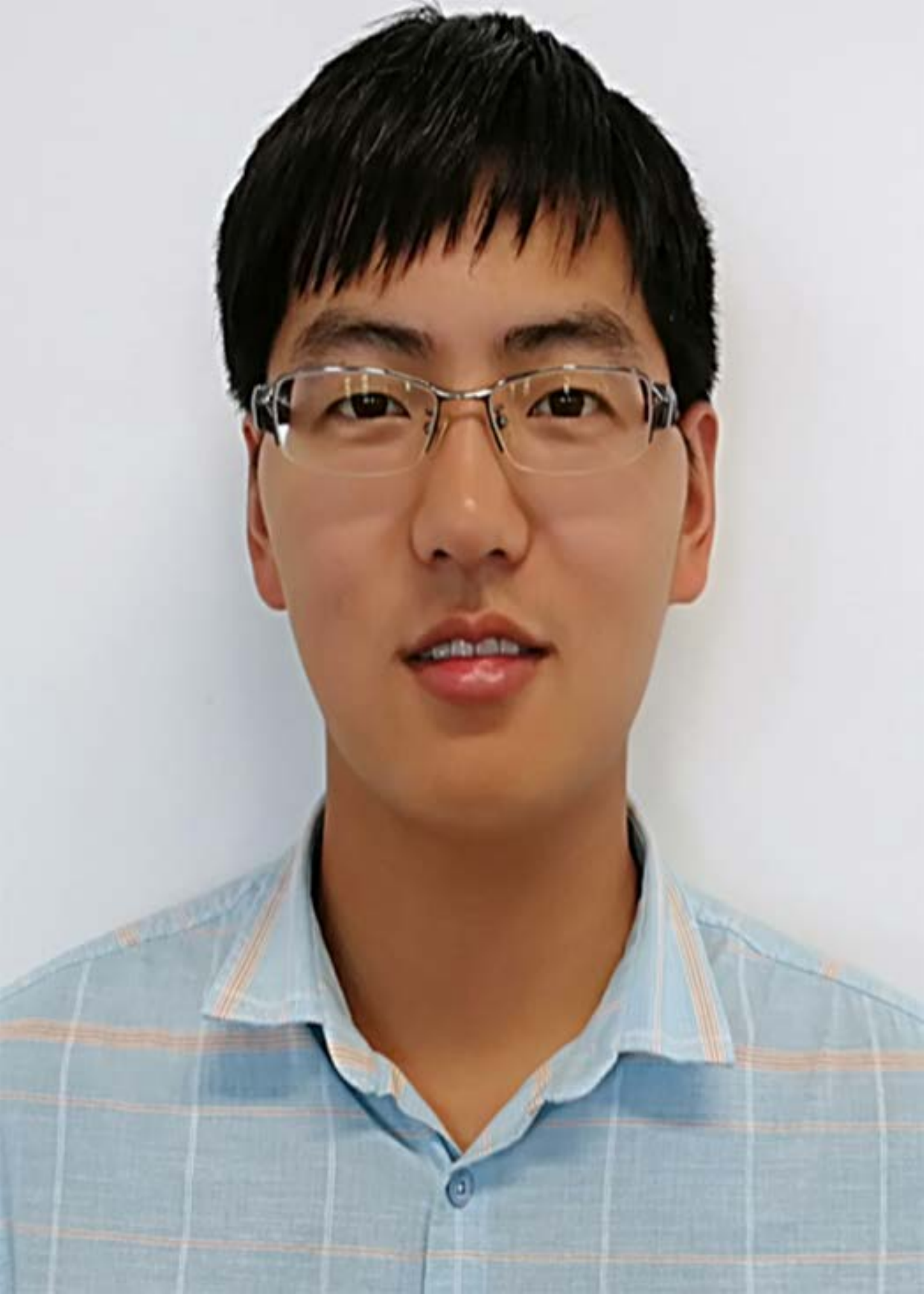}}]{Pengfei Jiao}
 received the Ph.D. degrees in computer science from Tianjin University, Tianjin, China, in 2018. From 2018 to 2021, he was a lecture with the Center of Biosafety Research and Strategy of Tianjin University. He is currently a Professor with the School of Cyberspace, Hangzhou Dianzi University, Hangzhou, China.
His current research interests include complex network analysis and its applications. 
\end{IEEEbiography}

\begin{IEEEbiography}[{\includegraphics[width=1in,height=1.25in,clip,keepaspectratio]{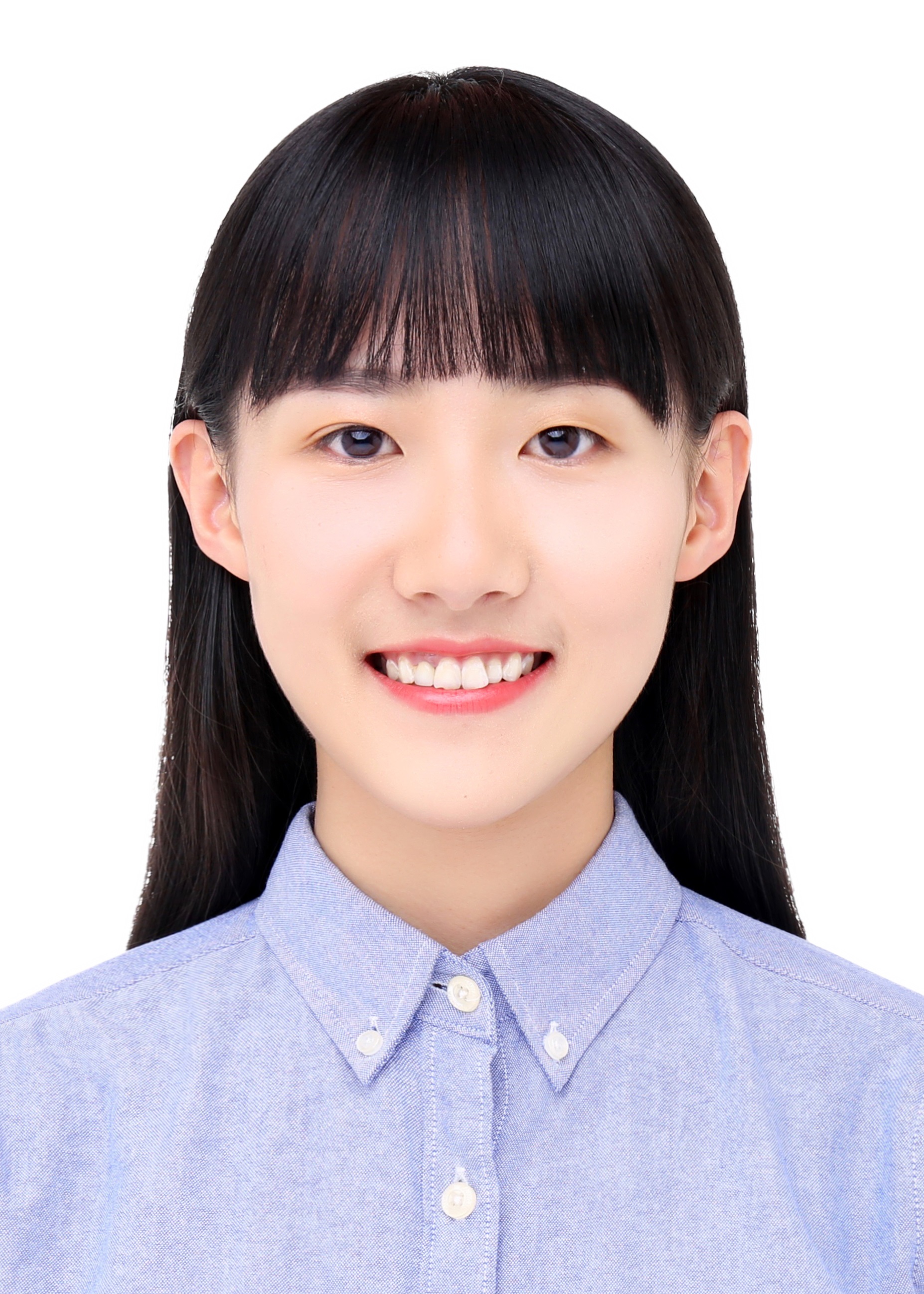}}]{Ting Pan} received the Bachelor degree from Xiamen University in 2020. She is currently pursuing a master's degree at the School of Computer Science and Technology, Tianjin University. Her current research interests include complex network analysis and role-based network representation learning.
\end{IEEEbiography}

\begin{IEEEbiography}[{\includegraphics[width=1in,height=1.25in,clip,keepaspectratio]{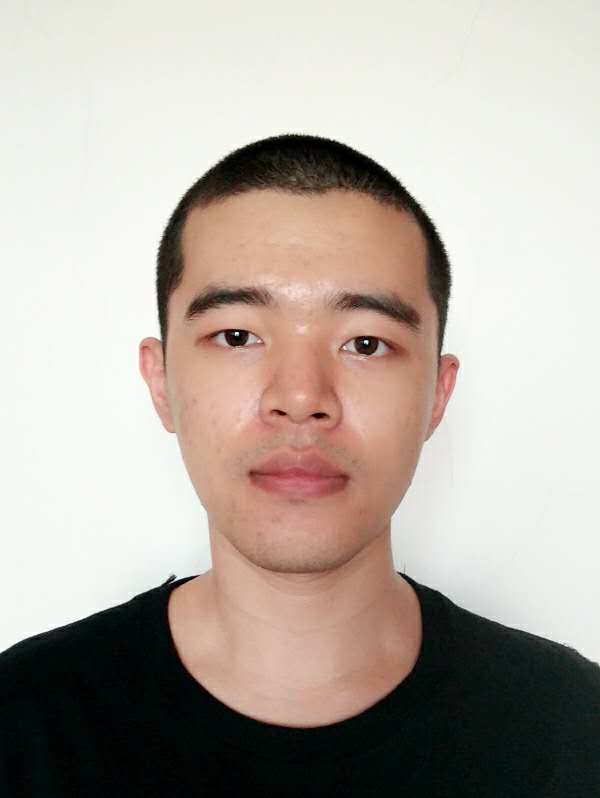}}]{Wang Zhang} received the Bachelor degree from Tianjin University in 2018. He is currently pursuing a master's degree at the School of Computer Science and Technology, Tianjin University. His current research interests include complex network analysis and network embedding.
\end{IEEEbiography}

\begin{IEEEbiography}[{\includegraphics[width=1in,height=1.25in,clip,keepaspectratio]{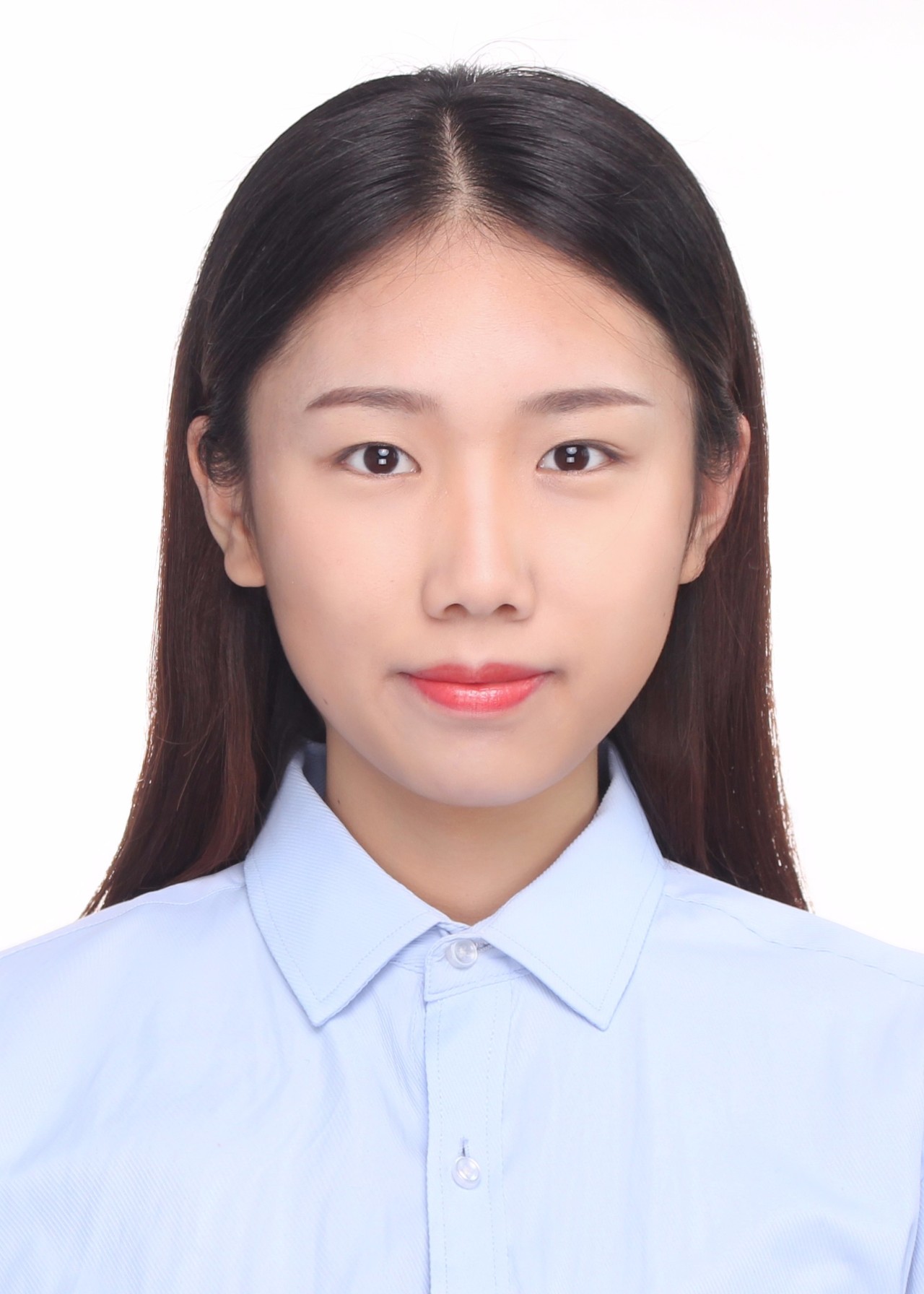}}]{Mengyu Jia}  received the Bachelor Degree in 2018 from Hangzhou Dianzi  University, Hangzhou, China. She is currently studying for a master's degree at the School of college of intelligence and computing, Tianjin University. Her research interests include complex network analysis and heterogeneou representaton learning.
\end{IEEEbiography}

\begin{IEEEbiography}[{\includegraphics[width=1in,height=1.25in,clip,keepaspectratio]{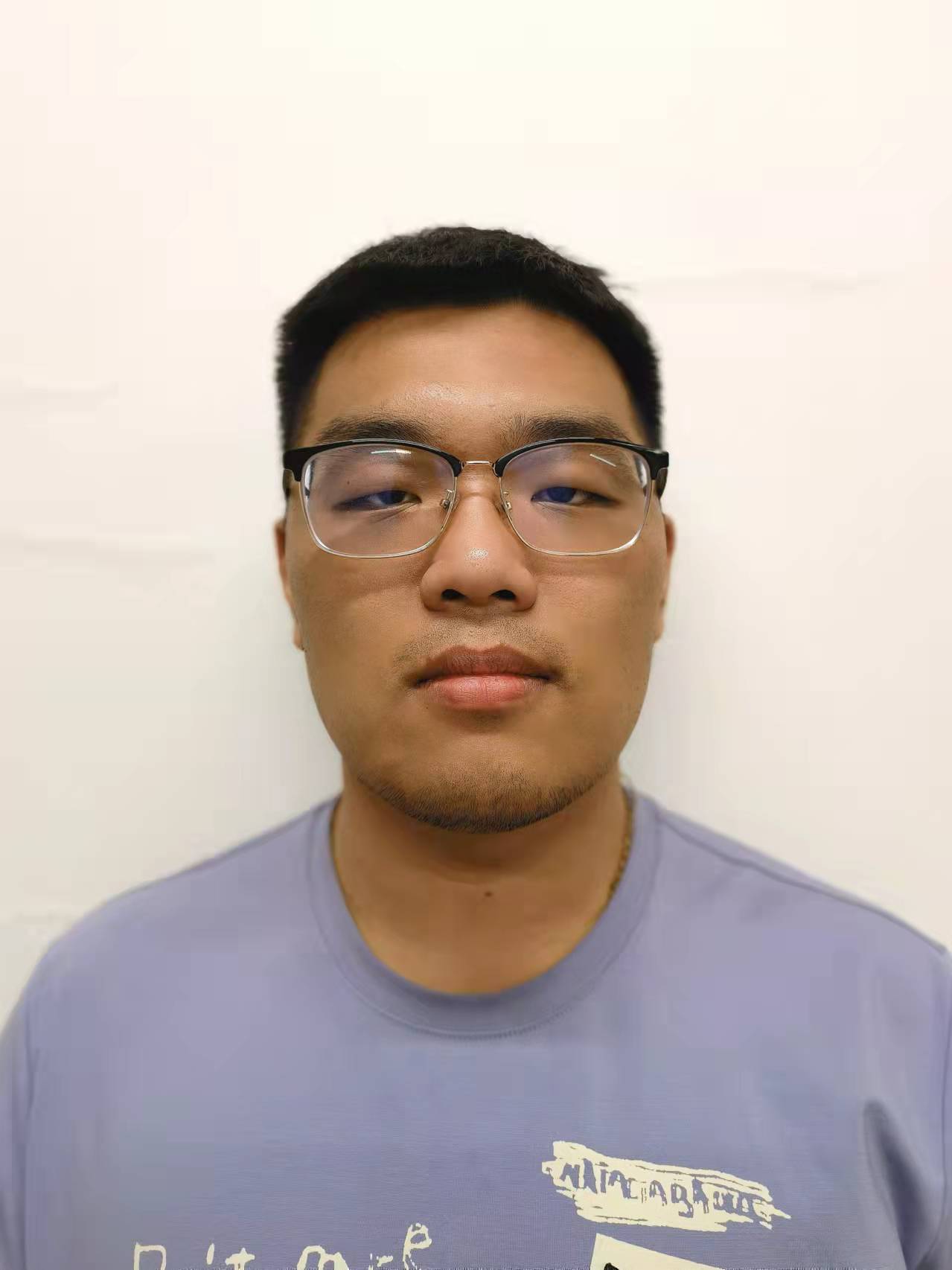}}]{Danyang Shi} received the Bachelor degree from Tianjin University in 2020.He is currently pursuing a master's degree at the College of Intelligence and Computing, Tianjin University. His current research interest is heterogeneous information network representation learning.
\end{IEEEbiography}

\begin{IEEEbiography}[{\includegraphics[width=1in,height=1.25in,clip,keepaspectratio]{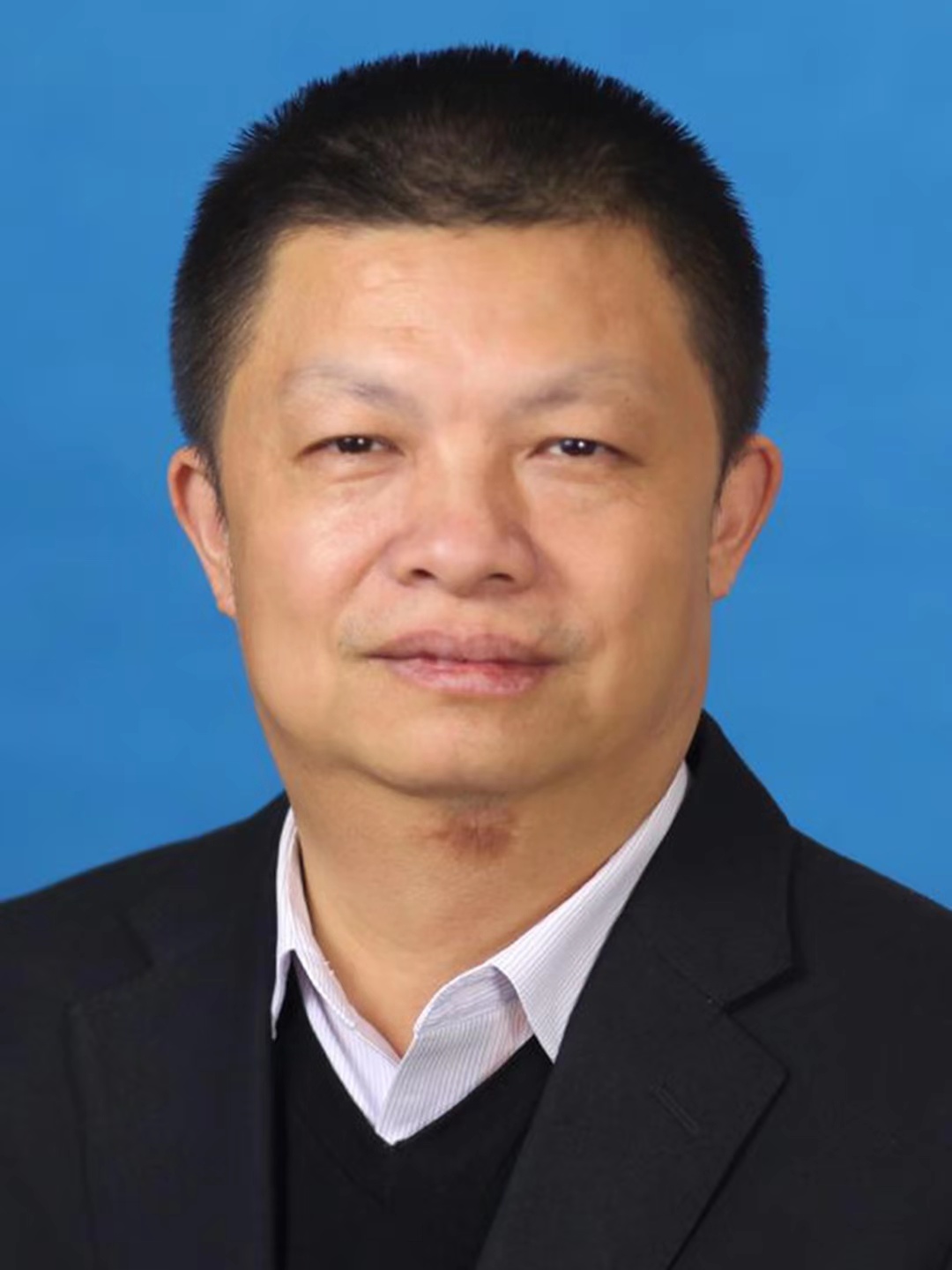}}]{Wenjun Wang} is currently a Professor at the School of College of Intelligence and Computing, Tianjin University, His research interests include computational social science, large-scale data mining, intelligence analysis and multi-layer complex network modeling. He has published more than 50 papers on main international journals and conferences.
\end{IEEEbiography}




\end{document}